\documentclass[preprint,12pt,authoryear]{elsarticle}

\newcommand\bcmdtab{\noindent\bgroup\tabcolsep=0pt%
 \begin{tabular}{@{}p{10pc}@{}p{20pc}@{}}}
\newcommand\ecmdtab{\end{tabular}\egroup}

\newcommand{\Prob}{\mathbb{P}}

\usepackage{algorithm}
\usepackage{algorithmic}
\usepackage{amsmath}
\usepackage{amssymb}
\usepackage{array}
\usepackage[english]{babel}
\usepackage{booktabs}
\usepackage{caption}
\usepackage{color}
\usepackage{epsfig}
\usepackage{etex}
\usepackage{gensymb}
\usepackage[a4paper]{geometry}
\usepackage{graphicx}
\usepackage{listings}
\usepackage{lineno}
\usepackage{mathptmx} 
\usepackage{multirow}
\usepackage{pst-all}
\usepackage{pst-node}
\usepackage{pst-plot}
\usepackage{pspicture}
\usepackage{pstricks}
\usepackage{subfigure}
\usepackage{theorem}
\usepackage{tikz}
\usepackage{url}
\usepackage[all]{xy}

\journal{Frontiers in Psychology: Quantitative Psychology and Measurement}

\begin{document}

\begin{frontmatter}



\title{Applying a Dynamical Systems Model and Network Theory to Major Depressive Disorder}


\author[label1]{Jolanda J. Kossakowski \corref{cor1}}
\author[label2]{Marijke C. M. Gordijn}
\author[label3]{Harri\"{e}tte Riese}
\author[label1]{Lourens J. Waldorp}

\address[label1]{Department of Psychology\\
		 University of Amsterdam\\
		 Nieuwe Achtergracht 129-B\\
		 1018 XE Amsterdam, The Netherlands\\}
\address[label2]{Department of Chronobiology\\
		 GeLifes, University of Groningen\\
		 Groningen, The Netherlands\\}
\address[label3]{Interdisciplinary Center Psychopathology and Emotion regulation\\ Department of Psychiatry\\ University Medical Center Groningen\\
		 University of Groningen, Groningen, The Netherlands}

\cortext[cor1]{Corresponding author}

\begin{abstract}
Mental disorders like major depressive disorder can be seen as complex dynamical systems. In this study we investigate the dynamic behaviour of individuals to see whether or not we can expect a transition to another mood state. We introduce a mean field model to a binomial process, where we reduce a dynamic multidimensional system (stochastic cellular automaton) to a one-dimensional system to analyse the dynamics. Using maximum likelihood estimation, we can estimate the parameter of interest which, in combination with a bifurcation diagram, reflects the expectancy that someone has to transition to another mood state. After validating the proposed method with simulated data, we apply this method to two empirical examples, where we show its use in a clinical sample consisting of patients diagnosed with major depressive disorder, and a general population sample. Results showed that the majority of the clinical sample was categorized as having an expectancy for a transition, while the majority of the general population sample did not have this expectancy. We conclude that the mean field model has great potential in assessing the expectancy for a transition between mood states. With some extensions it could, in the future, aid clinical therapists in the treatment of depressed patients.
\end{abstract}


\end{frontmatter}


\section{Introduction}

Major depressive disorder (MDD) is unfortunately not that uncommon: around 350 million people around the globe suffer from MDD \citep{WHO2012}. While many studies have been conducted in the treatment of MDD, it remains unclear why certain people develop MDD and others do not; we do not know the exact circumstances of the person and its environment that may lead to MDD. There is some empirical evidence that people experience discrete mood states \citep{Hosenfeld2015}. This has led to the hypothesis that mood changes or (sudden) transitions to MDD may be related to dynamical systems theory \citep{leemput2014, Cramer2016, Wichers2016}. In this paper, we build on these ideas to assess the expectancy that a person has to develop MDD and embed such assessments more thoroughly in dynamical systems theory and network theory in order to obtain a reasonable explanation of transitions to MDD. 

Recently, the idea has been put forward that mental disorders, like MDD, can be considered as a system of interacting variables \citep{Borsboom2011, Guloksuz:2017, Cramer2016, Kossakowski2018}. Aspects of MDD, like loss of energy or feelings of worthlessness, can be seen as nodes in a network that interact with, and influence each other at later times and other symptoms of MDD \citep{cramer2012}. This system of interacting emotions may change over time, making the system dynamic \citep{gulyas2013}. Connections between various aspects of MDD can increase or decrease in strength over time, or aspects themselves may increase or decrease in strength as an individual develops MDD.

We can measure these changes by means of the Experience Sampling Method \citep[ESM;][]{csikszentmihalyi1987}, where individual daily life experiences are measured several times a day for an extended period of time. At some point in time, when the system has surpassed some critical point \citep{Scheffer2014}, a discontinuous transition is made from a stable and healthy mood state to a stable and depressed mood state. These sudden jumps, called \emph{transitions} \citep{kuznetsov2013}, are central to complex dynamical systems, and are the subject of the assessment that we will undertake in the present paper.
 
Attempts to anticipate a transition are often approached by so-called \emph{early-warning signals} obtained from ESM studies \citep{Kossakowski2018}. Dynamical systems leave `breadcrumbs' behind in these time series that hint towards such a transition. These breadcrumbs occur before the transition, and after \emph{critical slowing down} that may occur when the system finds it more difficult to return to the original equilibrium state \citep{Scheffer2014}. Recently, it has been empirically shown that critical slowing down actually occurs prior to the transition \citep{leemput2014, Wichers2016}. While critical slowing down is an important line of research, it is difficult to analyse critical slowing down in a system that has more than a handful of variables.

\citet{Hosenfeld2015} introduced a statistical measure to determine whether there are one or two stable mood states, based just on the distribution of the number of active symptoms per measurement. This statistical measure, called the \emph{bimodality coefficient} (BC), only considers this distribution and determines whether there is evidence for one or two stable states. However, this approach offers no explanation of any kind of the phenomena observed in the distribution.

In this study we take a different approach and try to assess the expectancy of a transition  between mood states. We investigate this expectancy by combining dynamical systems theory with network theory. More specifically, we use cellular automata as the framework for networks (cellular automata) and their stochastic counterparts to investigate dynamic behaviour. There are three reasons why we believe that the dynamics of the stochastic cellular automaton may be appropriate for psychopathology. First, there is some evidence that mood states are discrete, or at least they are experienced as such (i.e., see \citealp{Hosenfeld2015}), and mood can switch between these states. A cellular automaton such as the one we propose is able to have multiple stable states that are discrete, and the process can `jump' between these states. The fact that the process can switch between states is important because we want to know the conditions under which such sudden changes can occur. Second, in line with network theory, we think that mood states and symptoms interact with each other and hence will influence each other (see \citealp{Borsboom2017}). A cellular automaton is a direct implementation of these ideas: it is a network and by definition each node affects its neighbours through an update rule, which can be specified based on the application. Third, because we always have uncertainty as to the correct specification of the variables in the network, we allow the updating process to be stochastic, accounting for unknown exogenous effects. 

We will simplify the automaton by reducing the network to a single dynamic equation (given certain assumptions), and by characterising the possible states of this reduced system. We then have a process that may be an accurate description of what is going on with the changes in symptoms over time. We can, in turn, analyse these changes analytically and through simulations. We assume (intuitively) that the nodes in the network function roughly in the same manner and that each of the nodes affects the others in a similar way. The assumptions lead to a so-called \emph{mean field model}. Using these assumptions, our focus becomes the proportion of active nodes in the system, which now forms a sequence of states ranging from 0 to 1. Since this sequence of states only depends on the proportion of active nodes at the previous time point, we obtain what is called a Markov chain and we can estimate the parameters by means of maximum likelihood estimation in a straight forward manner. Using this dynamical system allows us to determine whether for a person it is possible that a transition may occur or not. 


\begin{figure}
\centering
\includegraphics[width = 0.95\textwidth]{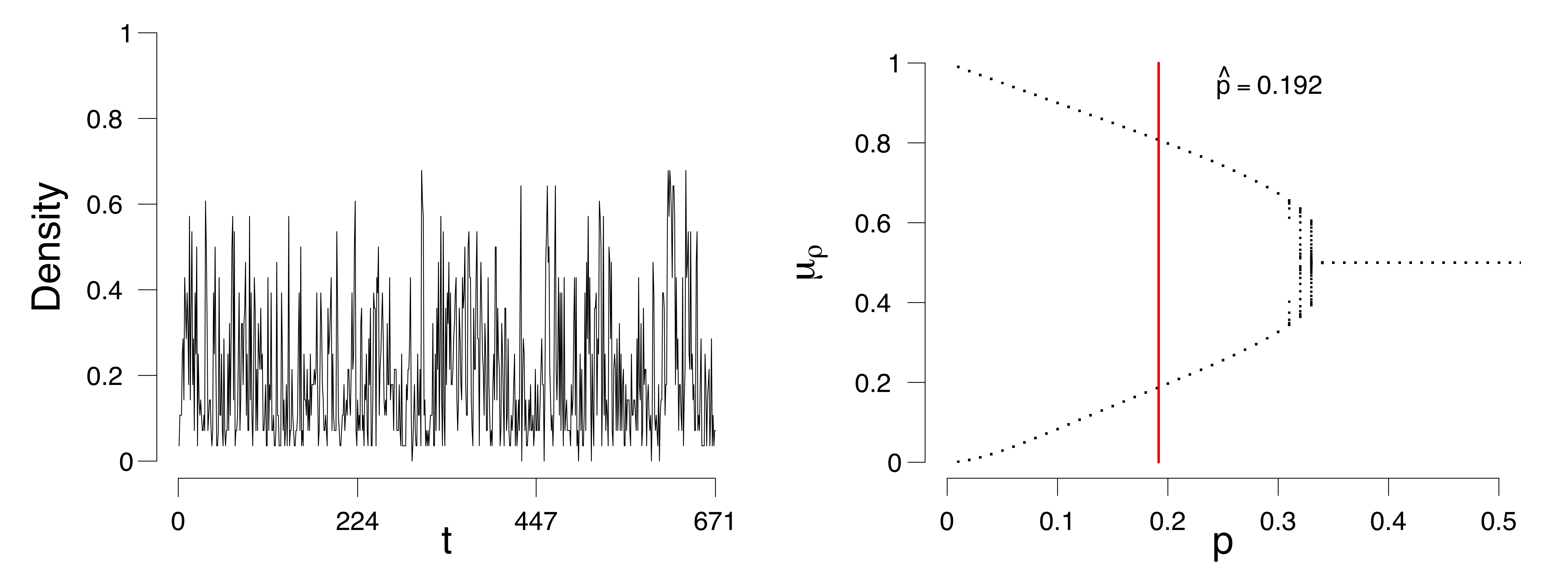}
\caption{The evolution of the percentage of active nodes for each time point (left figure), and the accompanying bifurcation diagram (right figure). The red line in the bifurcation diagram in the lower figures indicates the estimation of $p$.}
\label{fig:ESM_example}
\end{figure}

As an example, we consider a time series of the proportion of active emotions for a single subject, shown in Figure \ref{fig:ESM_example} (left). We identify the possible states of this person with respect to the network of emotions, depending on the parameter of the process we assume underlies these observations. For this process we can obtain a so-called \emph{bifurcation diagram} (Figure \ref{fig:ESM_example}, right). This bifurcation diagram shows the possible (likely) states for this person given a value on the probability $p$ of emotions changing from inactive to active. We assess from the time series of this person the parameters of our model and obtain an estimate of where in the bifurcation diagram this person is (represented by the vertical red line in Figure \ref{fig:ESM_example}, right). If the probability $p$ is in the range of $[0.34,0.50]$, where there is one point per value of $p$ on the $x$-axis, then this person will remain stable. If the probability is lower than approximately $0.34$, where there are two values for each value for $p$ on the $x$-axis, then there are two stable states, one with a high proportion of active emotions and one with a low proportion of active emotions. The estimate of the probability $p$ for this person is $0.192$ (the vertical red line in the right panel of Figure \ref{fig:ESM_example}). Based on this, we would classify this individual as someone who may expect a (sudden) increase in the proportion of active emotions and thereby experience an episode of depression. And indeed, for this individual we know (from external evidence) that a depressive episode had taken place after the time series that we used to determine the state of the person \citep[see][]{Wichers2016, Kossakowski2017}. 

In the present paper we obtain the maximum likelihood (ML) estimate for the model and the standard errors. We show, using simulations, that for many of the values of the parameter the estimate is reasonably close to the true value. Furthermore, we apply the proposed method to two real data sets, one with patients diagnosed with MDD, and one with subjects from the general population. This paper is set up as follows. First, we will briefly explain the theory of the mean field model and the proposed method. Then we present the simulation to show how the ML estimation performs. Finally, we apply our method to two datasets to show how the method works in different contexts.


%
\section{Stochastic cellular automata}
\begin{figure}
\centering
\includegraphics[width = 0.99\textwidth]{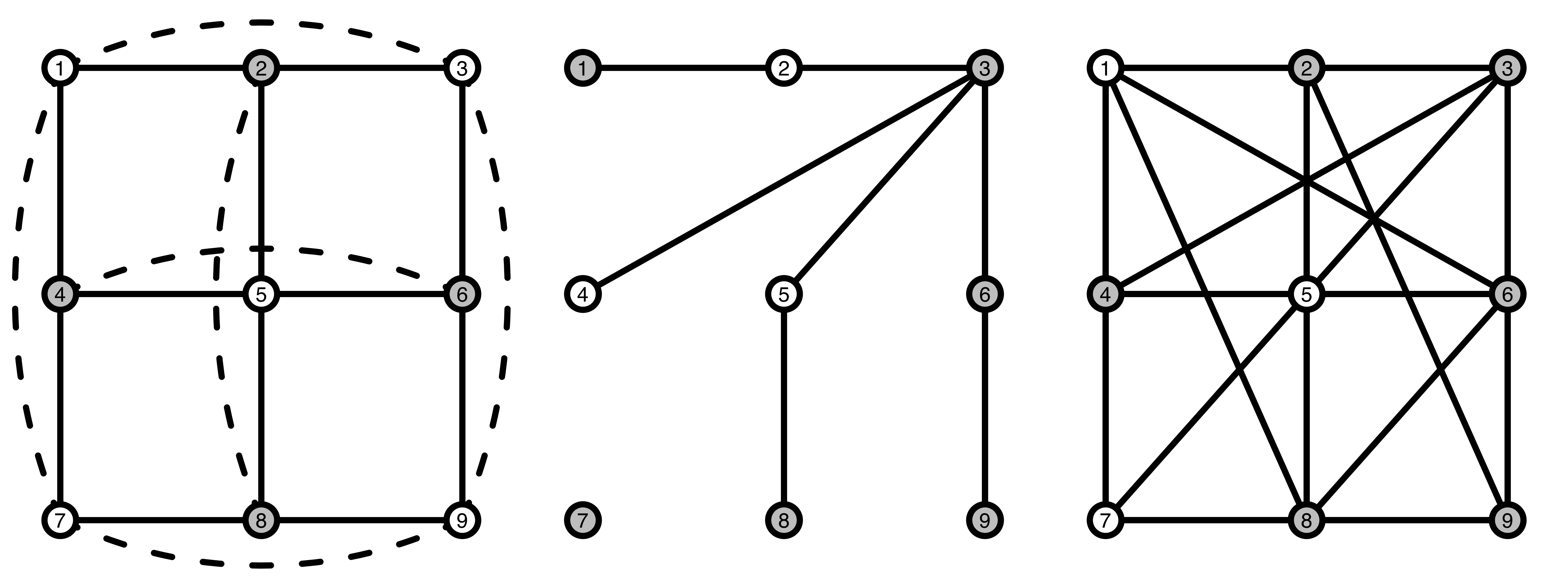}
\caption{Visualization of a grid structure (left figure), a random graph structure (middle figure), and a small-world structure (right figure). Grey nodes indicate the neighbours of the middle node in each graph. Solid lines indicate pairwise connections between nodes. Dashed lines are also pairwise connections, but have been curved and dashed as are hidden behind other connections in a 2D-view.}
\label{fig:torus}
\end{figure}

To model interacting symptoms and emotions we use a particular kind of structure, a \emph{stochastic cellular automaton} (SCA). Such automata are particular dynamical systems that show typical behaviour for stable and bistable behaviour depending on the settings \citep{kozma2004, Balister2006}, which is what we assume to the case for MDD. For the interested reader, the books by \citet{holmgren2012, hirsch2012, Hasselblatt2003, Golubitsky2003} provide background information on dynamical systems theory. A cellular automaton (CA) is a dynamical system where nodes are arranged in a fixed and finite grid, and where connected nodes determine the state of a node at each subsequent time point \citep{wolfram1984b, sarkar2000}. A node $j$ that is directly connected to node $i$ is called a neighbour. A grid is a graph $G_{\rm grid}(n,\Gamma)$ with $n$ nodes in the set $V=\{ 1,2,\ldots,n \}$ where each node $i$ has the same number of neighbours in its neighbourhood $\Gamma=\{j\in V:\text{$j$ {\rm is connected to} $i$}\}\cup \{i\}$ including itself. To ensure that all nodes have exactly the same number of neighbours, we impose the boundary condition such that a node at the boundary is connected to a node on the opposite end, making it a \emph{torus}. An example of such a grid is shown in Figure \ref{fig:torus} (left), where the center node is directly connected to its four neighbours, marked in grey. We consider elementary CAs where each node can be in either of two states: `active' (coded by $1$) or `inactive' (coded by $0$). In a CA a deterministic, local update rule $\phi$ determines the state $x_{i,t}$ of each node $i\in V$ at the next time step based on which nodes are active in the neighbourhood of node $x_{i,t}$. An example of such an update rule is the \emph{majority rule}, where each node becomes 1 (active) whenever more than half of the neighbours of node $i$ at the previous time point are active, and 0 (inactive) otherwise. Although many other update rules are possible, we will focus on this particular rule in the present study. One of the reasons for choosing the majority rule is that it is stable, i.e., for small changes in the number of active nodes the decision does not change \citep{ODonnell:2014}. Repeated application of the update rule $\phi$ results in a vector of 0s and 1s, called an orbit: At any time point $t$ the orbit $\phi^t(x_{i}) = \phi \circ \phi \circ \phi \cdots \circ \phi(x_{i,0})$ (initial value at $t = 0$), such that the same local rule is applied to the result of the previous time point $t$ times.

To illustrate, say that we have the network presented in Figure \ref{fig:torus} (left), and we have the following orbit of active and inactive nodes $\phi^0(x_{1,0}) = 1$, and for the other 8 nodes 1, 0, 1, 0, 0, 1, 1, 0, as shown in Table \ref{tab:major_example}. We can then determine how many active neighbours $r$ each node has, by just counting the number of active nodes each node is connected to. As mentioned in Table \ref{tab:major_example}, nodes 1, 5 and 7 have three active neighbours, nodes 2, 3, 4, 8 and 9 have two active neighbours, and node 6 has one active neighbour. For this example we will use the majority rule $\phi$ that is described earlier, which states that a node is activated (`$1$') when more than half of that node's neighbourhood is active. The majority rule uses $r>|\Gamma|/2$ to indicate whether the number of active neighbours is greater than half the size of the neighbourhood. $\Gamma$ here denotes the size of a node's neighbourhood. In our example, $|\Gamma| = 5$: each node has exactly four neighbours, and the node itself at $t-1$ is the fifth addition to $|\Gamma|$. With the majority rule $\phi$, the next time step becomes $\phi^1(x_{i,1}) = (1, 0, 0, 0, 1, 0, 1, 0, 0)$. We then use this sequence of active and inactive nodes to determine the number of active nodes $r$ at $t = 1$, which is described in Table \ref{tab:major_example}, column $r_1$. We can continue this process for a length $T$ (not shown in Table \ref{tab:major_example}), thus creating a $T \times n$ matrix that holds the orbit $\phi^t(x_{i,t})$ on the columns.

\begin{table}
\centering
\begin{tabular}{lllllll}
\hline
Node & $t_0$ & $r_0$ & $t_1$ & $r_1$ & $t_2$\\
\hline
1 & 1 & 3 & 1 & 1 & 0\\
2 & 1 & 2 & 0 & 2 & 0\\
3 & 0 & 2 & 0 & 1 & 0\\
4 & 1 & 2 & 0 & 3 & 1\\
5 & 0 & 3 & 1 & 0 & 0\\
6 & 0 & 1 & 0 & 1 & 0\\
7 & 1 & 3 & 1 & 1 & 0\\
8 & 1 & 2 & 0 & 2 & 0\\
9 & 0 & 2 & 0 & 1 & 0\\
\hline
\end{tabular}
\caption{Illustration of the majority rule as used for Figure \ref{fig:torus}. The columns $t_0$, $t_1$ and $t_2$ denote the sequence of active nodes at a specific time point. The columns $r_0$, $r_1$ and $r_2$ denote the number of active neighbours for per node at time $t$.}
\label{tab:major_example}
\end{table}

In the illustration above, the majority rule used to update the system was a deterministic one. In a stochastic cellular automaton (SCA), a probability is introduced to model uncertainty, based on the number of active neighbours ($r$). In our application to psychopathology, this uncertainty is required because we cannot predict the behaviour of emotions in our network exactly, and because we know that exogenous events influence these emotions that we cannot measure. By just counting the number of active neighbours that a node has, we can determine the probability for a node to become active. The probability $0\le p\le 1$ determines whether or not a node becomes active at time point $t+1$. The majority rule combined with this probability equals the probability that we obtain for node $x_{i,t+1}= 1$, given that there are $r$ active neighbours is

\begin{equation}
\mathbb{P}(X_{i, t+1} = 1\mid r)= \left \{
\begin{array}{ll}
p & \mbox{if } r \le |\Gamma|/2\\
1 - p & \mbox{if } r > |\Gamma|/2
\end{array}
\right.
\label{eq:psym} 
\end{equation}
where $|\Gamma|$ is the size of the neighbourhood and $r$ the number of active neighbours. The parameter $p$ is determined a priori or is estimated from data (see below). Because $\mathbb{P}(x_{i, t+1}\mid r)$ depends on the behaviour of the majority of a node's neighbourhood, this update rule is also called the \emph{majority rule}. In this SCA each node $i\in V$ is then updated according to the majority rule; all nodes are updated simultaneously (synchronous updating). The result for each node is a sequence (orbit) of 0s and 1s. From all $n=|V|$ nodes we can determine the total number of active nodes $Y_{t}$ at time point $t$, for all time points up to time $T$. We are interested in the number of active nodes $Y_t = \sum_{i=1}^n X_{i,t}$ (where $X_{i,t}$ is the value of node $x_i$ at time point $t$) and so we average over all nodes in the grid at each time point $t$, obtaining $\rho_{t}=Y_{t}/n$, which is often referred to in the literature as the \emph{density}. An example of the density (proportion) is shown in Figure \ref{fig:ESM_example} (left panel).


\begin{figure}
\includegraphics[width = 0.95\textwidth]{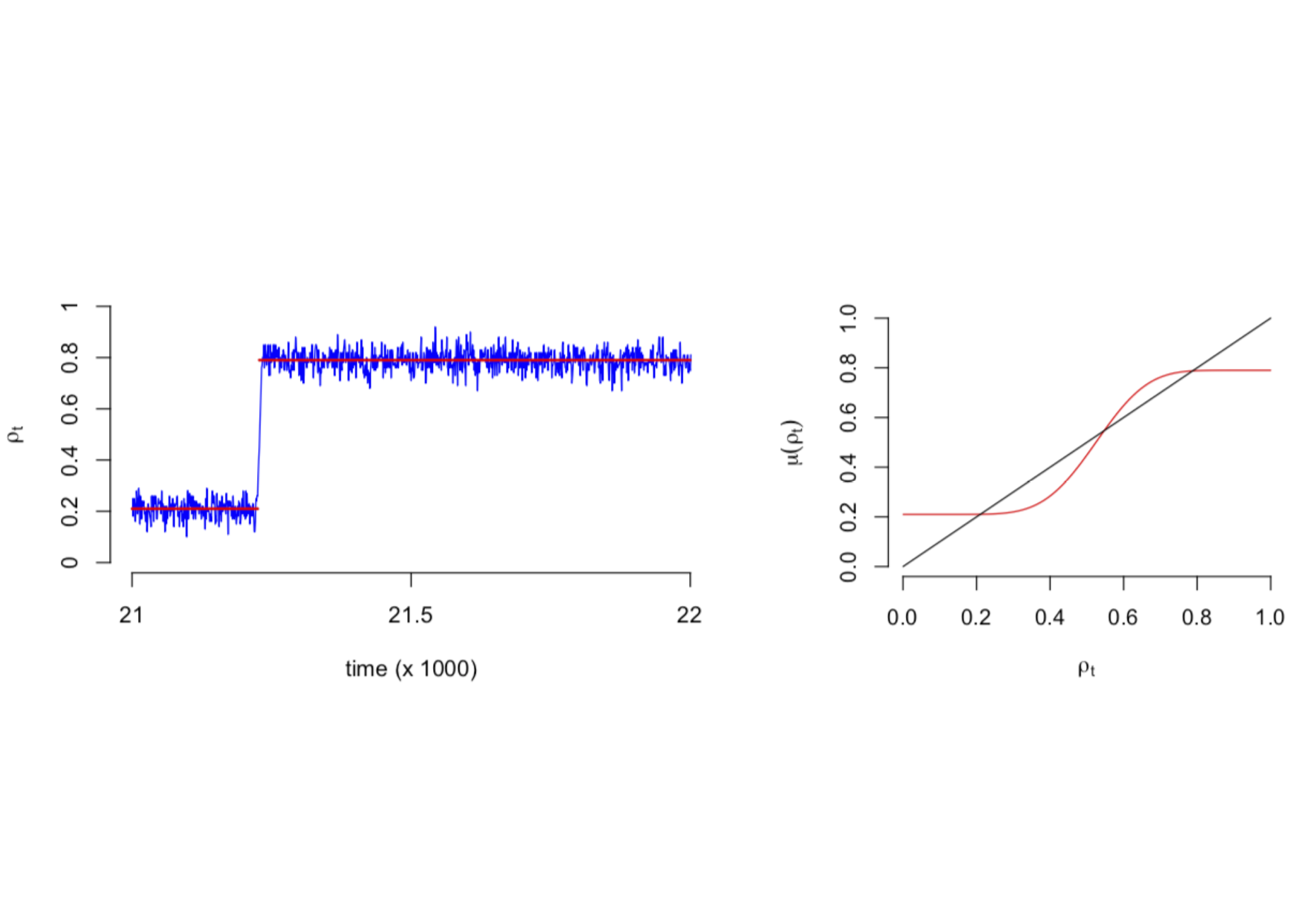}
\caption{Example of a stochastic cellular automata process that includes a transition (left). Example of the corresponding mean field function. The red line indicates the expectation of equation (\ref{eq:mfatorus}) divided by the number of nodes.}
\label{fig:example_process}
\end{figure}

\section{Mean field model}

It is rather difficulty to infer the characteristics of what the system will do in the long run from an SCA \citep{Lebowitz1990}. We need to simplify the SCA in order to make it possible to derive the characteristics of the SCA. Here we use an approximation for the structure of the network, where we assume the average of the number of neighbours for each node $|\Gamma|$. We also assume that nodes can be in either of two states: active (`1') or inactive (`0'), and that the nodes behave in a similar manner. The latter assumption means that the majority rule, as presented in equation (\ref{eq:psym}), is applied to all nodes in the network, and that all nodes become active or inactive in the same way. In the grid (Figure \ref{fig:torus}, left) it is easily seen that each node is similar to any other node since each node has the same number of neighbours, and becomes active or inactive in the same way by means of the majority rule that is based on the number of active neighbours. This allows us to simplify an SCA to a single discrete time dynamical system, as in \citet{kozma2005}, \citet{Balister2006}, and \citet{waldorp2019}. 

In the {\em mean field} model we make use of the uniformity of the nodes in a grid. The change of a node from $0$ to $1$ or vice versa is based on the number of active neighbours in that node's neighbourhoor ($r$) and the probability parameter $p$, following the majority rule defined in (\ref{eq:psym}). In a grid each node has exactly the same number of neighbours and so the probability of a node changing value depends on the properties of the grid and not on the local activity. Therefore, as shown in \citet{kozma2005} and \citet{Balister2006}, we obtain at time point $t+1$ the number of active nodes in the grid $Y_{t+1}$, which is a random variable with a binomial distribution that has a success probability $\rho_{t}=Y_{t}/n$, the proportion of active nodes (density) at the previous time point $t$. The number of draws in the binomial probability is determined by the size of the neighbourhood $|\Gamma|$ particular to the graph. The majority rule in (\ref{eq:psym}) determines for which number of active nodes we obtain $p$ up until active nodes $r\le \lfloor |\Gamma|/2\rfloor$, where $\lfloor a \rfloor$ is the integer part of $a$, or $1-p$ otherwise. To define the probability of the number of neighbours that are 1, we need to consider all possible configurations of $|\Gamma|$ active-inactive nodes in the graph. There are $\binom{|\Gamma|}{r}$ ways to choose $r$ active nodes out of $|\Gamma|$ each with a success probability $\rho_{t}=Y_{t}/n$. We then obtain for the probability of $r$ active nodes out of $|\Gamma|$ 
\begin{equation}
\mathbb{P}(r\mid \rho_{t})={|\Gamma| \choose r}\rho_t^r(1-\rho_t)^{|\Gamma|-r} 
\label{eq:mf-base}
\end{equation}
Simultaneously, we require the probability $p$ or $1-p$ from the majority rule in (\ref{eq:psym}), which is assumed to be independent. We need to define the probability for any number of active nodes and therefore marginalise over the number of possible active nodes in the neighbourhood $r$. Putting (\ref{eq:psym}) and (\ref{eq:mf-base}) together, we obtain the joint probability $\mathbb{P}(X_{i, t+1} = 1 \mid r, \rho_t) = \mathbb{P}(X_{i,t+1}=1 \mid r) \mathbb{P}(X_{i,t+1} \mid \rho_t)$. Hence we obtain the probability for any node in the graph to be 1 as

\begin{equation}
\rho_{t+1}^{\rm grid} = \sum^{|\Gamma |}_{r=0} \mathbb{P}(X_{i, t+1} = 1 \mid r){|\Gamma| \choose r}\rho_t^r(1-\rho_t)^{|\Gamma|-r} 
\label{eq:mf-base-full}
\end{equation}
%
%
%

Because the evolution is binomial based on the proportion of active nodes at the previous time point (see equation (\ref{eq:mf-base-full})), it follows from the transition probability that the number of active nodes $X_{t+1}=x_{t+1}$, given that at $t$ is $X_{t}=x_{t}$ in the graph $G_{\rm grid}$, is 
\begin{align}\label{eq:transition-prob}
\Prob(X_{t+1}=x_{t+1}\mid X_{t}=x_{t})= \binom{n}{x_{t+1}}\rho_{t}^{\rm grid}(x_{t}/n)^{x_{t+1}}(1-\rho_{t}^{\rm grid}(x_{t}/n))^{n-x_{t+1}}
\end{align}
So, we know how in a grid with $n$ nodes the proportion of active nodes $\rho_t$ changes from time point $t$ to time point $t+1$, for any $t$. The mean field model uses the mean of this binomial process and divides by $n$ to obtain the proportion. We often denote the expected value of $Y_{t+1}/n$ by $\mu_{\rm grid}:=\rho_{t}^{\rm grid}$ to emphasise that we use the mean of the process in a grid. We know that the fluctuations around the mean are small (depending on the standard deviation and size of the grid, see \citealt{waldorp2019}), so the mean is a good approximation of the process itself.

As an illustration of the binomial process, in the left panel of Figure \ref{fig:example_process} we see a typical SCA process where it is clear that the fluctuations are around a particular mean ($0.2$) for time points before $t = 21000$ approximately. After this point (tipping point) the fluctuations revolve around another mean ($0.8$) with a higher proportion of emotions. In the right panel of Figure \ref{fig:example_process} we see a plot of the expectation of the process, which is the mean field that predicts the values at which the mean of the process converges to eventually. It is this mean function that we will use to represent the process and the network that evolves over time.

We now regard the mean field, the expectation of the binomial process $E(Y_t/n) = \mu_{\rm grid}$, as the dynamical system that is a representation of the network. This dynamical system evolves by repeated application of $\mu_{\rm grid}$ to its previous result. We analyse the dynamical properties of $\mu_{\rm grid}$ by considering a so-called bifurcation diagram. Plugging in different values for the a priori parameter $p$ from (\ref{eq:mf-base-full}) in the majority rule, in the interval $(0,0.5]$, we obtain a \emph{bifurcation diagram}, as shown in Figure \ref{fig:ESM_example} (right panel). In a bifurcation diagram the repeated application of $\mu_{\rm grid}$ is applied to updated values of $\rho_{t}^{\rm grid}$ such that the last section of the orbit is displayed where the process is in equilibrium \citep[stable if stable fixed points exist;][]{hirsch2012}. For each value of $p$, displayed on the $x$-axis in Figure \ref{fig:ESM_example}, one sequence is generated, of which the last 50 are displayed in Figure \ref{fig:ESM_example}. In some cases, the sequence will find two equilibria, and thus we draw two points at those two equilibria. In other cases, the sequence will converge to one equilibrium, and thus only one point will be drawn in the bifurcation diagram. Such diagrams show what kind of behaviour can be expected to be generated by the process. Here we see that there are two kinds of situations: (a) a stable situation when $p$ is in the interval $(0.34,0.50]$, where irrespective of the starting point, the process ends up at that stable fixed point, and (b) a bistable situation when $p$ is in $[0,0.34]$ where the process could (suddenly) switch between states (transition) to a low or high density. The parameter value $p$ at which the process changes from a stable to a bistable situation is called the critical point. In Figure \ref{fig:ESM_example} the critical point lies at $p \approx 0.34$; the parameter area $0< p \le 0.34$ is bimodal where transitions can occur, whereas the parameter area $0.34<p<0.50$ represents a unimodal area where the mean field is stable. Thus, the parameter $p$ can be used to determine whether a process has two stable states, and therefore can transition between them, or one stable state, where no transition can occur.

The probability for the mean field in (\ref{eq:mf-base-full}) is designed for a grid with a fixed neighbourhood size $|\Gamma|$. In the context of psychology and psychopathology, it is hard to come up with a graph representing the interactions between variables, that would take the form of a grid. We therefore also looked at the mean field model for a \emph{random graph} and a \emph{small-world graph}. A random graph $G_{\rm rg}(n,p(e))$ is a graph structure with $n$ nodes and a (constant) probability $p(e)$ for an edge to be present in the graph \citep{bollobas2001, durrett2007}. In the mean field model of a random graph, the neighbourhood size $|\Gamma|$ is a random variable that is maximally $n-1$. Each node has a binomial number of neighbours with expected number of neighbours $p(e)(n-1)$. We extend the idea used for the grid, where we marginalise over all possible configurations of number of active nodes for each neighbourhood of size $n-1$ for the random graph. One can approximate this probability accurately with a small modification of the probability used for a grid \citep{waldorp2019}. The difference with the probability on the grid is in the size of the neighbourhood (see Figure \ref{fig:torus}, left and middle panel), where in the grid the neighbourhood size is fixed to $|\Gamma|$. In the mean field model for the random graph, we fix this to the expected number of neighbours $p(e)(n-1)$. Let $\nu=\lfloor p(e)(n-1)\rfloor$ be the integer part of the expected number of neighbours. For the random graph $G_{\rm rg}$ the neighbourhood size is no longer $|\Gamma|$ (like it is for the grid), but $\nu$. The probability in a random graph for a node to become active given the graph's density at time point $t$ ($\rho_t$) and the edge probability then becomes \citep{waldorp2019}:

\begin{equation}
\rho_{t+1}^{\rm rg} = \sum^{\nu}_{r=0} \mathbb{P}(X_{i, t+1} = 1 \mid r){\nu \choose r}\rho_t^r(1-\rho_t)^{\nu-r} 
\label{eq:mfarandomgraph}
\end{equation}

A small-world graph is in between a grid and a random graph where, compared to a random graph, the average clustering is high and the average path length is low \citep{watts1998}. A modified version of the small world is the Newman-Watts (NW) small-world \citep{newman1999}. In the NW small-world $G_{\rm sw}(n,\Gamma,p(w))$ the $n$ nodes each have a neighbourhood $\Gamma$ as in the grid and edges are added to the network following a (constant) wiring probability $p(w)$ \citep{newman1999}. We can then split up the probability in a part associated with the grid and a part associated with the random graph. The part for the grid is adjusted such that it corresponds to no other edges being present, i.e., we obtain $\rho_{t}^{\rm grid}(1-p(w))^{n-|\Gamma|}$, where the product $(1-p(w))^{n-|\Gamma|}$ represents the probability that no other edges are present for $n-|\Gamma|$ nodes. For the random part we obtain the probability as in (\ref{eq:mfarandomgraph}) but the first $|\Gamma|$ edges left out, because they have already been accounted for by the grid part. We denote this probability by $\rho_{t}^{\rm rg,\Gamma}$, which denotes the probability as in (\ref{eq:mfarandomgraph}) but with $\rho_{t}^{\rm rg,\Gamma}$ starting at $|\Gamma|+1$ instead of $0$. Then the probability for a node to become active given the graph's density at time point $t$ ($\rho_t$) and the wiring probability in the small-world graph $G_{\rm sw}$ is

\begin{equation}
\rho_{t+1}^{\rm sw} = \frac{|\Gamma|}{n}\rho_{t}^{\rm grid}(1-p(w))^{n-|\Gamma|} + \frac{n-|\Gamma|}{n}\rho_{t}^{\rm rg,\Gamma}
\label{eq:mfasmallworld}
\end{equation}
The small-world probability is therefore a combination of the probability on the grid and on a random graph, proportionately weighted. 


\section{Estimation of probability $p$ and graph parameters}


Our objective is to derive an estimate of the probability parameter $p$ from a time series to determine whether an individual can expect a transition between two mood states. One way of obtaining such an assessment is to determine where in a bifurcation diagram a person is located with respect to the parameter $p$ in the majority rule; is this in the stable area, where no transition can occur, or is it in the bistable area where a transition can occur. In order to do this, we need to estimate the parameter $p$ that is essential in the majority rule in (\ref{eq:psym}). Here we use maximum likelihood (ML) to obtain an estimate of $p$ \citep{Rajarshi2012}. 

If we take a closer look at equation (\ref{eq:mf-base-full}), it can be noticed that all parameters are known prior to the analysis, with the exception of the probability parameter $p$. To obtain $p$, we can estimate it from the data using ML estimation. We then obtain the maximum of the log-likelihood for the probability parameter $p$ that exists in (\ref{eq:psym}). We write the transition probability in going from state $x_t$ to state $x_{t+1}$ (number of active nodes in the graph) in (\ref{eq:transition-prob}) from $t$ to $t+1$ as $\Prob(X_{t+1}=x_{t+1}\mid X_{t}=x_{t})$. The log of the joint probability function (loglikelihood) for the number of active nodes is then 
\begin{equation}\label{eq:loglik}
\log \mathbb{P}(X_{t},t\ge 0) = \sum_{t=0}^{T-1} \log \Prob(X_{t+1}=x_{t+1}\mid X_{t}=x_{t})
\end{equation}

\noindent where $T$ denotes the total duration of the sequence in time points. The transition probability $\mathbb{P}$ is as in (\ref{eq:transition-prob}). The data that are plugged into this equation is a vector of length $T$ that holds the number of active nodes for each time point $t$. At each time point the number of active nodes is given as input to the probability in the binomial process $\rho_t = (Y_t/n)$, where $x_t$ is the number of observed active nodes at time $t$. The data are plugged in the transition probabilities, where we recognise in the SCA that we can relatively easily find the transition probability to go from $x_{t}$ to $x_{t+1}$ active nodes. We can find these transition probabilities because of the fact that we have, for each of the graphs $G_{\rm grid}$, $G_{\rm rg}$, and $G_{\rm sw}$, a binomial process with a probability of success particular to each type of graph. The parameter $\rho_t$ for the random graph $G_{\rm rg}$ and the small-world graph $G_{\rm sw}$ are similar except that we change the probability of success to $\rho_{t}^{\rm rg}$ or $\rho_{t}^{\rm sw}$, respectively. 
 
The process is ergodic whenever the probability $\rho_{t}^{\rm grid}$ is not in the basin of attraction of 0 and 1 \citep[see][]{waldorp2019}. In other words, a process is ergodic when the process is stationary and when all nodes in the graph follow the same dynamics \citep{Molenaar2007}. In those cases we could simplify expression (\ref{eq:loglik}) using only the transition probabilities that do not depend on time. In general, however, we do not know where the probabilities are, and therefore we do not assume ergodicity and cannot simplify the log-likelihood to terms consisting only of the states and not on time \citep{Fleming1978}. We maximise the log-likelihood function in (\ref{eq:loglik}) with respect to $p$ to obtain its estimate from an empirical time series, making it possible to place that person on the bifurcation diagram and assess the expectancy of possible switching. 

\begin{figure}
\centering
\includegraphics[width = 0.99\textwidth]{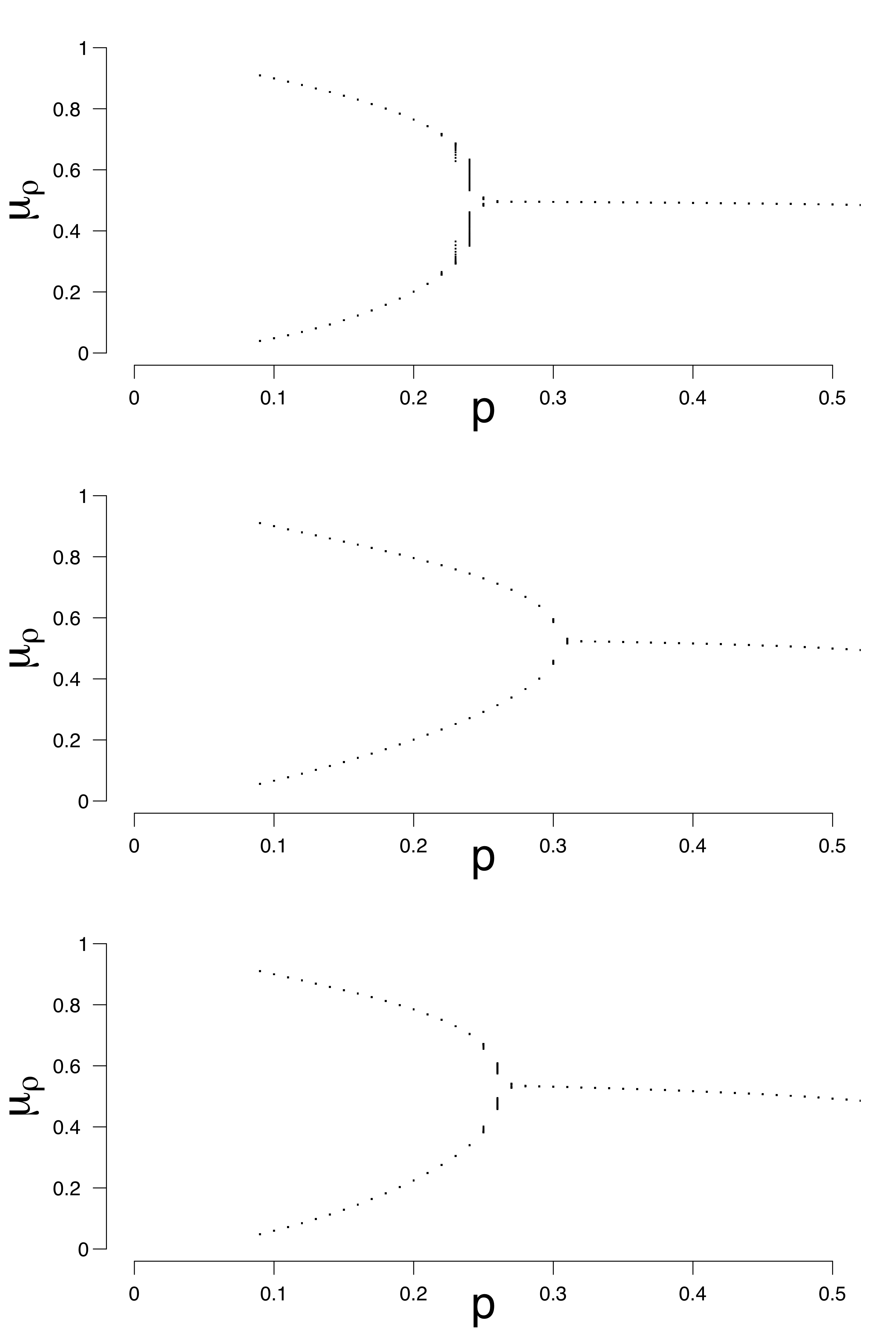}\\
\caption{Examples of bifurcation diagrams for a grid (top figure), a random graph (middle figure; $p(e) = 0.1$) and a small-world graph (lower figure; $p(w) = 0.1$).}
\label{fig:bifurcation_examples}
\end{figure}

In both the random and small-world graph we have additional graph parameters: in the random graph we have the probability of an edge $p(e)$, and in the small-world graph we have the probability of re-wiring $p(w)$. Both parameters are obtained by maximising the log-likelihood with respect to $p(e)$ and $p(w)$ respectively. Equations (\ref{eq:mf-base-full}), (\ref{eq:mfarandomgraph}) and (\ref{eq:mfasmallworld}) each show how we calculate the density ($\rho$) in a grid, a random graph, or a small-world graph, respectively. One only needs to plug in a value for $p$ (and the graph parameters $p(e)$ or $p(w)$ in the case of a random graph or a small-world graph, respectively) into the equation and let it run for some time $T$ (often 1000 is enough), to find out at what density it will end up, or between which two values it may transition in the case of two stable states. By varying the value for $p$, one can create a bifurcation diagram, of which examples are shown in Figure \ref{fig:bifurcation_examples}. Each dot represents a separate run of the mean field equation. Equation (\ref{eq:mf-base-full}) is reflected in the top panel of Figure \ref{fig:bifurcation_examples}, (\ref{eq:mfarandomgraph}) in the middel panel, and (\ref{eq:mfasmallworld}) in the bottom panel of Figure \ref{fig:bifurcation_examples}.

Taking the top panel of Figure \ref{fig:bifurcation_examples} as an example, if we run equation (\ref{eq:mf-base-full}) with $p = 0.1$, it can be seen that the binomial process ends up in either $0.1$ or $0.9$ approximately, and could switch between these states. Our mean field model says that if we estimate the probability parameter $p$ for an individual to be $\hat{p} \approx 0.1$, then this person could experience a transition between the two states, which could be related to a depressive episode. When we increase the value of the probability parameter $\hat{p} \approx 0.3$, the binomial process no longer has the possibility of a transition between states, but will remain around $0.5$ approximately. The critical point, the point where the system changes from having two stable states to one stable state, differs depending on the size of the graph and the type of graph; for the random graph and the small-world graph the location of the critical point also depends on the graph parameters $p(e)$ or $p(w)$, respectively, as seen in Figure \ref{fig:bifurcation_examples}. To summarise, in order to categorise individuals, we need to know the individual's position in terms of the probability parameter $p$ in its personalised mean field model.

Uncertainty can be quantified by the standard error of the estimate $\hat{p}$. For the grid we have only the estimate of $p$ and for the random graph and the small-world we have the edge probability $p(e)$ and $p(w)$, respectively. Standard errors are obtained from the second order derivatives (Hessian) of the log-likelihood \citep{Rajarshi2012}. The inverse (matrix) of the Hessian and scaled by $1/T$ results in the variance of the parameter estimate. The square root of the diagonal elements are the standard errors, i.e., $\text{SE}(\hat{p}) = \sqrt{\frac{1}{T}h^{11}}$ is the standard error for $\hat{p}, \text{SE}(\hat{p}(e)) = \sqrt{\frac{1}{T}h^{22}}$ is the standard error for the edge probability in the random graph, and $\text{SE}(\hat{p}(w)) = \sqrt{\frac{1}{T}h^{22}}$ is the standard error for the rewiring probability, where $h^{ij}$ is the $ij$th element of the inverse Hessian. 

\section{Validation of probability $p$ and graph parameters}

Before we apply the mean field model to empirical data, we want to know how well the mean field model can estimate the probability parameter $p$ in simulated data. We simulated 100 networks for each topology of a grid, a random graph, and a small-world graph. We varied the size of the network $V$ $\in$ $\left\{\right.16, $ $25,$ $49,$ $100\left.\right\}$, the number of time points $T$ $\in$ $\left\{\right.50,$ $100,$ $200,$ $500,$ $5000\left.\right\}$, and the probability $p$ $\in$ $\left\{\right.0.1,$ $0.2,$ $0.3,$ $0.4,$ $0.5,$ $0.6,$ $0.7,$ $0.8,$ $0.9\left.\right\}$ (Equation \ref{eq:mf-base-full}). We also varied the probability for an edge in the random graph $p(e)$ $\in$ $\left\{\right.0.1,$ $0.2,$ $0.3,$ $0.4,$ $0.5,$ $0.6,$ $0.7,$ $0.8,$ $0.9\left.\right\}$ (Equation \ref{eq:mfarandomgraph}) and the probability for an edge to be rewired in the small-world graph $p(w)$ $\in$ $\left\{\right.0.1,$ $0.2,$ $0.3,$ $0.4,$ $0.5,$ $0.6,$ $0.7,$ $0.8,$ $0.9\left.\right\}$ (Equation \ref{eq:mfasmallworld}). For $t = 0$, a random number of nodes was set to active by using the {\tt\small R} package \emph{IsingSampler} version 0.2 \citep{IsingSampler}. 

For each of the 100 simulation runs, we used the $T \times n$ set of active and inactive nodes to estimate the probability parameter $p$ and the graph parameters $p(e)$ and $p(w)$. All simulated data, figures, and the used R-code are publicly available \citep[OSF;][]{OSFKosII}. For clarity of presentation, figures are only presented for $T=50$, as results for the other number of time points were nearly identical. We also only present the results for $p$, $p(e)$ and $p(w) \in \{0.1, 0.2, 0.3, 0.4, 0.5\}$ as the simulation results for these parameters $> 0.5$ hardly occur in empirical data, and are therefore for this paper less interesting. These and other results can be found online \citep{OSFKosII}. For each simulation run, we calculated the absolute difference between the probability parameter $p$, under which the data were simulated, and $\hat{p}$, which we estimated from the data using ML estimation. We denoted this absolute difference by ${\Delta}(p)$, after which we take its mean ($\overline{\Delta}(p)$). This mean is determined for each replication, and can be interpreted as an error rate. The lower this value, the closer the estimate $\hat{p}$ is to the original value $p$. The same procedure was performed to determine the accuracy for graph parameters $p(e)$ ($\overline{\Delta}(p(e))$) and $p(w)$ ($\overline{\Delta}(p(w))$). A complete overview of all results across all conditions can be found in Table S1 in the supplementary files.

Figure \ref{fig:3D-accuracy-p} shows a visual representation of the mean absolute difference ($\overline{\Delta}(p)$) between the true probability parameter $p$, and the estimated probability parameter $\hat{p}$. It can be seen that the error rate $\overline{\Delta}(p)$ for $p$ is low for all different network structures. Supplementary file S1 shows the mean estimate of $p$ and its associated standard deviation for all conditions. The standard deviation for $\hat{p}$ is pretty low across conditions and never exceeds $0.04$. The mean error rate $\overline{\Delta}(p)$ did not exceed $0.08$ for the grid (for $T = 5000$, $n = 100$, $p = 0.4$), $0.06$ for the random graph (for $T = 50$, $n = 25$, $p = 0.2$, $p(e) = 0.1$), and $0.04$ for the small world graph (for $T = 50$, $n = 16$, $p = 0.5$, $p(w) = 0.4$). The error rate ranged between $0.006 - 0.12$ for the grid, $0.0009 - 0.15$ for the random graph, and $0.008 - 0.16$ for the small world graph. A small increase in the error rate can be noticed for the grid around the values $p = 0.3$ and $p = 0.4$. We think that a possible explanation is that the mean field model has some issues with estimating $p$ around the critical value, the point where the system either has one stable state, or two stable states. Because of fluctuations in the process, the exact critical point is difficult to estimate.

\begin{figure}
\centering
\includegraphics[width = 0.99\textwidth]{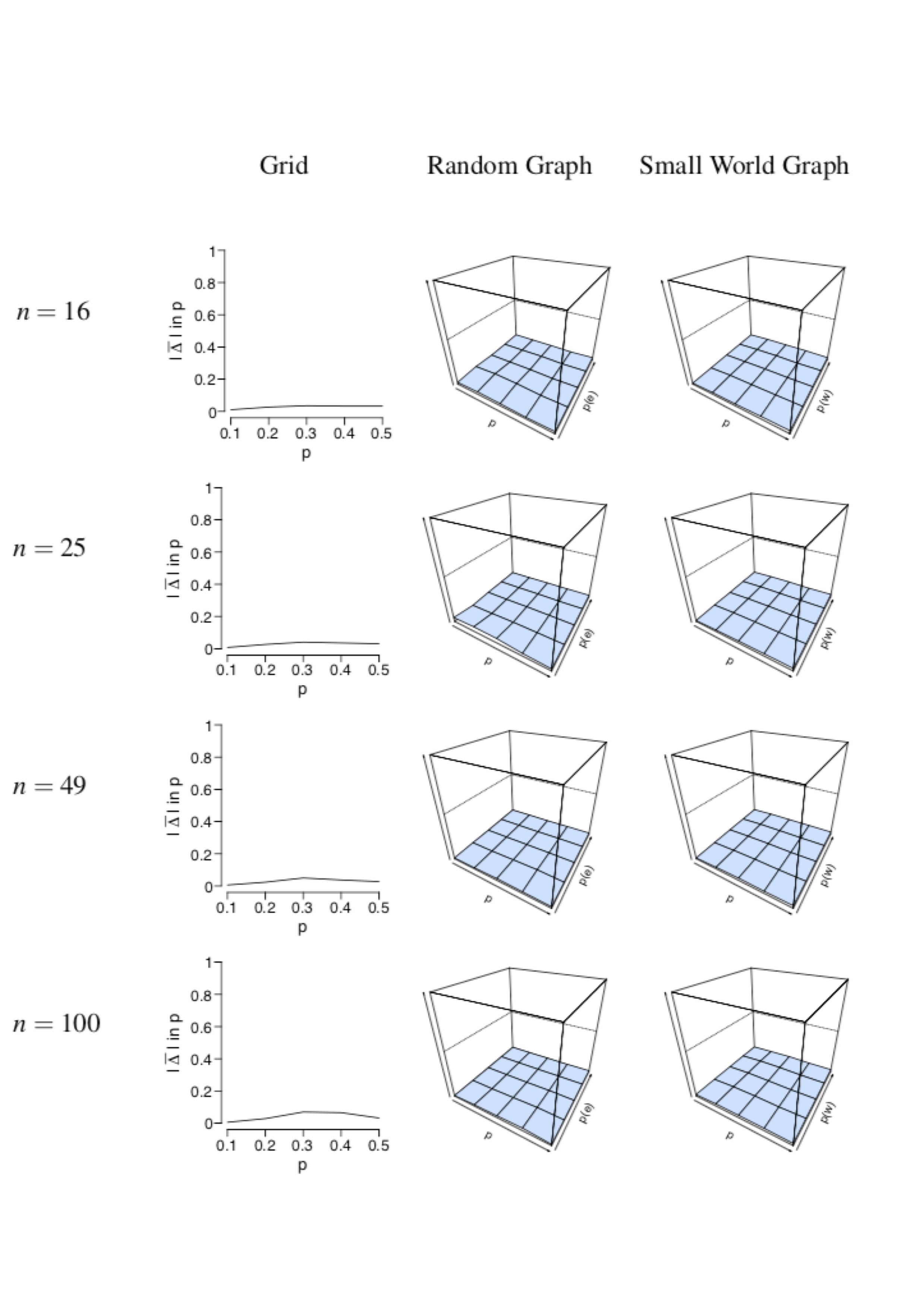}\\
\caption{Visualization of the mean error rate between $p$ and $\hat{p}$. Mean absolute difference is shown for a grid (left column), a random graph (middle column), and a small world graph (right column) at $T = 50$. For the left column, the x-axis denotes the parameter $p$ for which we simulated data, and the y-axis the mean absolute difference between $p$ and $\hat{p}$. For the middle and right column, the x-axis denotes the parameter $p$ for which we simulated data, the z-axis the graph parameter that was used to simulate data, and the y-axis the mean absolute difference between $p$ and $\hat{p}$ and runs between $0$ and $1$.}
\label{fig:3D-accuracy-p}
\end{figure}

\begin{figure}
\centering
\includegraphics[width = 0.99\textwidth]{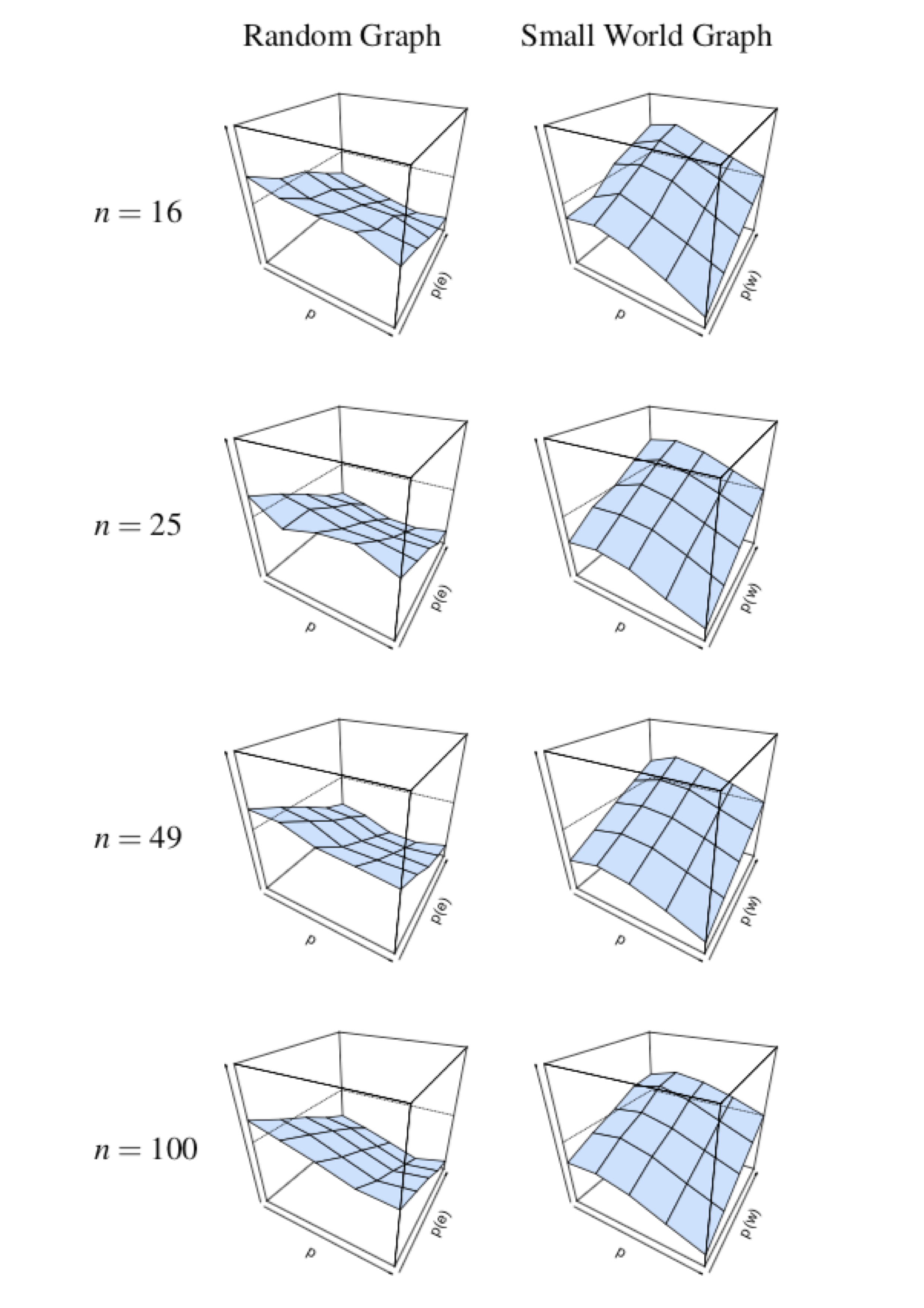}\\
\caption{Visualization of the mean absolute differences between $p(e)$ and $\hat{p}(e)$ and $p(w)$ and $\hat{p}(w)$ at $T=50$. Mean absolute difference is shown for a random graph (left column) and a small world graph (right column). The x-axis denotes the parameter $p$ for which we simulated data, the z-axis the graph parameter that was used to simulate data, and the y-axis the mean absolute difference between $p$ and $\hat{p}$, and runs between $0$ and $1$.}
\label{fig:3D-accuracy-nspc}
\end{figure}

The same conclusion cannot be drawn for graph parameters $p(e)$ and $p(w)$, as seen in Figure \ref{fig:3D-accuracy-nspc}. For a random graph, $\overline{\Delta}(p(e))$ is high when $p(e)$ is low, and decreases as $p(e)$ is increased. This shows that the graph parameter $\hat{p}(e)$ is most accurate when $p(e)$ is high. A possible explanation for this finding could be found in the connectedness of random graphs. When $p(e)$ is small, the probability that not all nodes are connected increases, resulting in isolated nodes. When we look at the minimum probability $p(e)$, such that the graph is connected for different network sizes, we see that $p(e)$ must be at least $0.46$ when the network size is $16$, $0.31$ when the network size is $25$, $0.17$ when the network size is $49$ and $0.09$ when the network size is $100$. Thus, as $p(e)$ increases, the probability for the network to be connected increases, and as a result of this, the error $\overline{\Delta}(p(e))$ decreases. The reverse is true for a small-world graph, where $\overline{\Delta}(p(w))$ is high when $p(w)$ is high and $p$ is low, and that it decreases when $p(w)$ also decreases. This shows that the graph parameter $\hat{p}(w)$ is most accurate when $p(w)$ is low and when $p$ is high. 

%

To investigate the standard errors, we calculated the mean standard error (SE) and its associated standard deviation for all conditions using the Hessian matrix provided by the ML estimation procedure. Table S1 in the supplementary files depict the mean SE and its standard deviation for all conditions. It can be seen that the mean SE is extremely low across all conditions, indicating good accuracy of the estimates. We calculated the absolute difference between the standard deviation of $\hat{p}$ and the SE of $\hat{p}$. The difference ranged from $0.0003$ to $0.18$, and in $98.9\%$ of all conditions, the difference between the standard deviation and the SE is smaller than $0.05$.

\begin{figure}
\centering
\includegraphics[width = 1\textwidth]{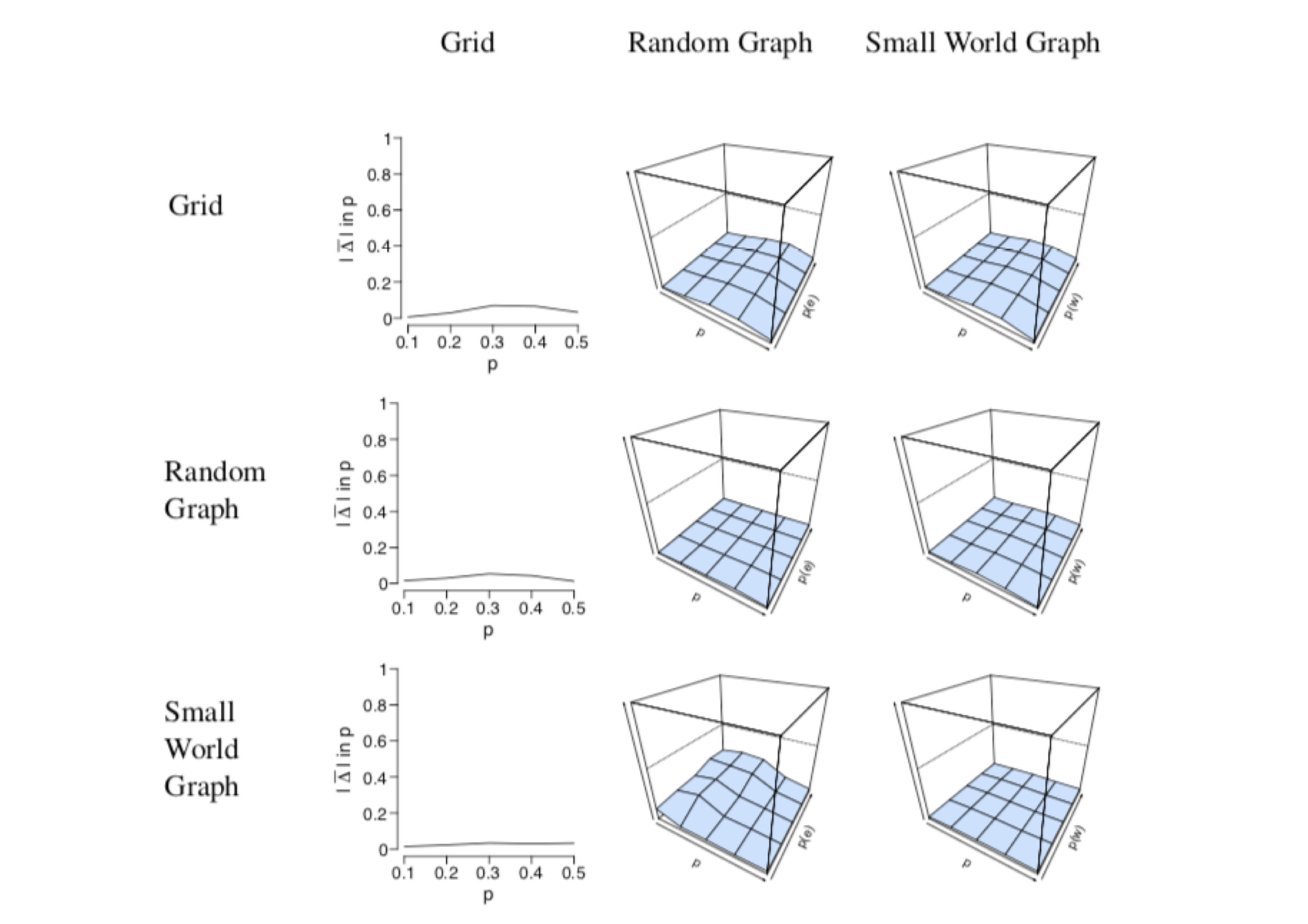}\\
\caption{Visualization of the mean absolute differences between $p$ and $\hat{p}$ that resulted from the misspecification analysis with $n = 100$ nodes and $T = 50$ time points. The rows denote the structure for which the data were simulated. The columns denote the structure for which $\hat{p}$ was estimated. $p$ for which we simulated data, and the y-axis the mean absolute difference between $p$ and $\hat{p}$. For the middle and right column, the x-axis denotes the parameter $p$ for which we simulated data, the z-axis the graph parameter that was used to simulate data, and the y-axis the mean absolute difference between $p$ and $\hat{p}$ that ranges between $0$ and $1$.}
\label{fig:misspec_results}
\end{figure}

Next to the SE, we calculated the error rate $\Delta(p)$ when the network structure is misspecified, and thus used the incorrect model to estimate $\hat{p}$ from the data. We used two datasets that represent the best and worst case scenario in terms of data structure. The worst case is the data with $n = 100$ nodes and $T = 50$ time points. The best case is the data with $n = 16$ nodes and $T = 5000$ time points. By taking the least and most ideal combination of $n$ nodes and $T$ time points, we obtain results where all other combinations will most likely lie in. With these properties in mind, we selected the data simulated for all three network structures, and applied all three models to estimate $\hat{p}$. Figure \ref{fig:misspec_results} depicts the error rate $\overline{\Delta}(p)$ for $n = 100$ nodes and $T = 50$ time points. We chose not to present the results for the data with $n = 16$ nodes and $T = 5000$ nodes for clarity of presentation. We estimated $\hat{p}$ for all three network structures for all datasets. It can be seen in Figure \ref{fig:misspec_results} that $\overline{\Delta}(p)$ is generally low, regardless of the network structure that was used to simulate the data.

For each estimation we calculated the \emph{Bayesian Information Criterion} (BIC) and compared it to the BIC of the other two network structures. The BIC is used for model selection, where the model with the lowest BIC is most preferred \citep{Wit2012}. Results showed that the grid structure was never the preferred network structure. The random graph is often selected ($63.8\%$ of the cases across conditions) as the preferred network structure when the data are simulated under a random graph. The small-world graph is preferred over the random graph at $p(e) = 0.1$ or $p(e) = 0.2$. A possible explanation for this is that, at this value for $p(e)$, the network is very sparse and it may be difficult to distinguish between the network structures. For data simulated under the small-world graph, the small-world graph itself is most often selected based ($69.3\%$ of the cases across conditions) on the BIC. There are no conditions in which the random graph is preferred over the small-world graph. It is worthy to note that, as $p$ increases, the difference in mean BIC between the network structures decreases, and more often the ``incorrect'' model is selected. This is also shown in Figure \ref{fig:bifurcation_examples}, where there is little difference between the bifurcation diagrams, especially when $p$ is high.

\begin{figure}
\centering
\includegraphics[width = 1\textwidth]{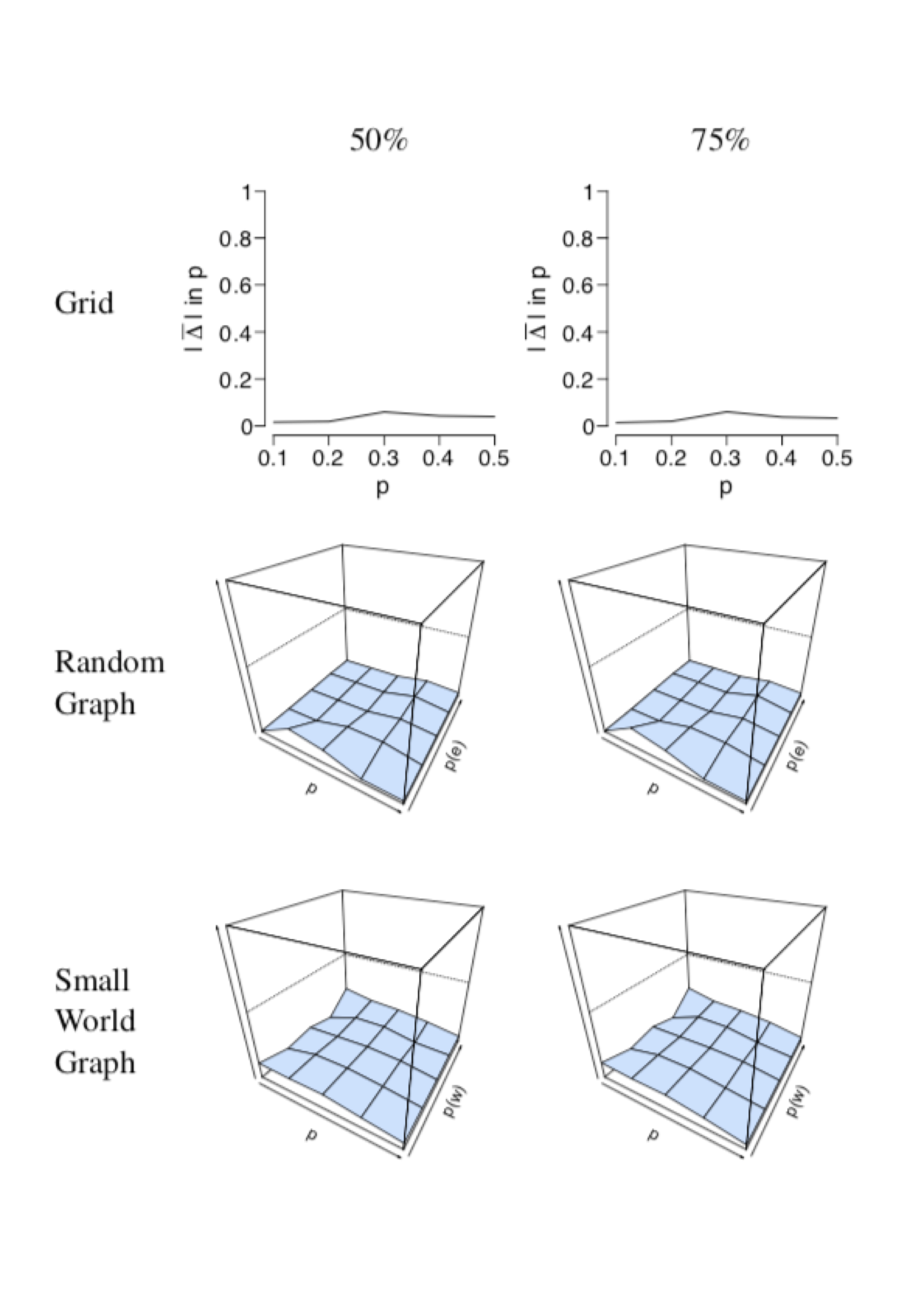}\\
\caption{Visualization of the mean absolute differences between $p$ and $\hat{p}$ that resulted from the subset analysis with $n = 100$ nodes and $T = 50$ time points. The rows denote the structure under which the data were simulated and analysed. The left columns shows the result for the subset analysis with $50\%$ of the data retained, while the right column shows the results with $75\%$ of the data retained. The x-axis denotes the parameter $p$ for which we simulated data, the z-axis the graph parameter that was used to simulate data (in case of the 3D figures), and the y-axis the mean absolute difference between $p$ and $\hat{p}$.}
\label{fig:subset_accuracy_p}
\end{figure}

As a last measure to study the robustness of our ML estimation, we performed a subset analysis, taking either $50\%$ or $75\%$ of the simulated time points to estimate $\hat{p}$. Similar to the misspecification analysis that we described previously, we looked at data with $n = 100$ nodes and $T = 50$ time points, and data with $n = 16$ nodes and $T = 5000$ time points. For each simulation condition, we randomly selected one simulation and selected a subset of the data, which we repeated 100 times. Figure \ref{fig:subset_accuracy_p} shows the mean error rate $\overline{\Delta}(p)$ between $p$ and $\hat{p}$ for $n = 100$ nodes and $T = 50$ time points. It can be seen that the mean error rate is generally low for all conditions and network structures. This means that, even when we take a subset of the data, the mean field model is able to correctly estimate $p$ from the data that we used.

%

In sum, the mean field model estimates $p$ well from the data; the graph parameters $p(e)$ and $p(w)$ could not be estimated as accurately. For the random graph parameter $p(e)$, this could potentially be solved by taking the ratio of edges present in the graph, and the total number edges possible in the graph. Alas, there is no similar solution for the small-world graph parameter $p(w)$. In the application of the mean field model, we assume all graphs to be random graphs. As estimating $p(e)$ from the data and extracting it from the graph resulted in nearly identical results, we decided to use the former option.

\section{Application to empirical time-series data}

Here, we will demonstrate how the probability $p$ of an emotion to be active is estimated from empirical data. In the following sections, we will show two empirical examples and demonstrate how the proposed method works in each of these examples. By showing the application of our proposed method on two different kinds of data, we aim to show how our proposed method works for different participants, and different types of data. The first example is a dataset of patients who were admitted as patients to a closed, psychiatric ward of an academic hospital \citep{Gordijn1994, Gordijn1998}. The second example is a dataset of healthy participants who were originally recruited in a nation-wide study \citep{VanderKrieke2015}. 

The data in these examples are time-series data. When collecting these types of data, participants are asked to complete a questionnaire several times a day. These questionnaires often contain items regarding a participant's current mood state, but can also hold items regarding a participant's physical condition, for example. In both examples, participants received a `beep' on fixed times during the day and were asked to complete the questionnaire. These beeps, in turn, correspond to the time points in time-series data. For example, when a participant completed twenty questionnaires, the data contains $T = 20$ time points. All analyses were performed using the R statistical software 3.4.4 \citep{RCoreTeam2016}. 

Next to the estimation of the probability parameter $p$, we calculate the \emph{Bimodality Coefficient} \citep[BC;][]{Hosenfeld2015} for each participant in both datasets, and compare the outcome of the two measures. The BC only takes information from the distribution of the proportion of active nodes (density) to determine whether there is evidence for one or two stable states. The BC is calculated as follows:
\begin{equation}
BC = \frac{s^2 + 1}{k + C}
\end{equation}
where $s$ is the skewness of the distribution, $k$ the kurtosis of the distribution, and $C$ a correction factor that depends on the number of variables: $C = \frac{3(n-1)^2}{(n-2)(n-3)}$. The BC obtains values between 0 and 1 and considers values $>0.55$ to mean there is evidence for two states \citep{Hosenfeld2015}. We only use the BC for comparison, the BC uses no specific information or assumptions about the process, only distributional properties are involved. We have no reason to believe that the BC and our proposed method MFA should or should not correspond since there is no clear connection between the two measures.

\subsection{Example 1: Clinical sample}

This example involves a secondary analysis of data that were originally gathered for a study in patients diagnosed with MDD, who were admitted to a closed, psychiatric ward of an academic hospital \citep{Gordijn1994, Gordijn1998}. The data have been described in detail in previous papers \citep{Gordijn1994, Gordijn1998}. Patients in this study completed the Dutch version of the Adjective Mood Scale \citep[AMS;][]{VonZerssen1986} twice a day at fixed time points for a period of six weeks, resulting in a maximum of 84 measurements per patient. Patients had to indicate on this 28-item questionnaire which of two given emotions (or neither) corresponded most closely to the patient's emotion at that moment in time. A detailed description of the items of the AMS can be found in Table \ref{tab:Groningen_labels}.

\begin{table}
\centering
\begin{tabular}{p{0.5cm}p{0.75cm}p{5.5cm}p{0.75cm}p{4cm}p{2cm}}
\hline
 & More & Dutch/English & More & Dutch/English & Neither/nor \\
\hline
1 & 0 & Openhartig/openly & 2 & Geremd/inhibited & 1 \\
2 & 0 & Welgemoed/good mood & 2 & Droefgeestig/bad mood & 1 \\
3 & 2 & Inactief/passive & 0 & Bedrijvig/active & 1 \\
4 & 2 & Ziekelijk/sickly & 0 & Kiplekker/healthy & 1 \\
5 & 0 & Doelbewust/purposefully & 2 & Doelloos/aimlessly & 1 \\
6 & 2 & Ernstig/serious & 0 & Geestig/humorous & 1 \\
7 & 2 & Fantasieloos/unimaginative & 0 & Fantasierijk/imaginative & 1 \\
8 & 0 & Gevoelig/sensitive & 2 & Gevoelloos/numb & 1 \\
9 & 2 & Pessimistisch/pessimistic & 0 & Optimistisch/optimistic & 1 \\
10 & 0 & Zorgeloos/carefree & 2 & Tobberig/worried & 1 \\
11 & 2 & Gebroken/broken & 0 & Monter/cheerful & 1 \\
12 & 0 & Liefderijk/lovingly & 2 & Liefdeloos/loveless & 1 \\
13 & 2 & Schuldig/guilty & 0 & Onschuldig/innocent & 1 \\
14 & 2 & Uitgeput/tired & 0 & Uitgerust/rested & 1 \\
15 & 2 & Levensmoe/life-tired & 0 & Levenslustig/lively & 1 \\
16 & 0 & Goed/good & 2 & Slecht/bad & 1 \\
17 & 0 & Vrolijk/cheerful & 2 & Treurig/tearful & 1 \\
18 & 0 & Bemind/loved & 2 & Onbemind/unloved & 1 \\
19 & 2 & Lui/lacking in energy & 0 & Actief/energetic & 1 \\
20 & 2 & Gesloten/withdrawn & 0 & Open/sociable & 1 \\
21 & 0 & Levendig/lively & 2 & Levenloos/sluggish & 1 \\
22 & 0 & Temperamentvol/temperamentfull & 2 & Futloos/lifeless & 1 \\
23 & 0 & Oplettend/watchful & 2 & Verstrooid/absent & 1 \\
24 & 2 & Wanhopig/desperate & 0 & Hoopvol/hopeful & 1 \\
25 & 0 & Tevreden/satisfied & 2 & Ontevreden/dissatisfied & 1 \\
26 & 2 & Angstig/anxious & 0 & Strijdlustig/combative & 1 \\
27 & 0 & Krachtig/powerful & 2 & Krachteloos/powerless & 1 \\
28 & 0 & Evenwichtig/balanced & 2 & Gejaagd/agitated & 1 \\
\hline
\end{tabular}
\caption{Items of the Adjective Mood Scale (AMS) and their assigned labels. Items marked with a * have a reversed response scale. The English translation may differ from the original AMS scale, as well as the order of te items.}
\label{tab:Groningen_labels}
\end{table}




We dichotomized the data by collapsing the `neither' condition with the positive mood state per individual item. We coded the positive mood state as `0' and the negative mood state as `1'. We also collapsed the `neither' condition with the negative mood state and ran the analyses with these data, but as these results were very similar to the ones we present, we left it out of this study. After dichotomizing the data, we replaced any missing measurements with the previous measurement. We also considered removing the missing measurements entirely, but as we found nearly identical results, we chose not to present these results. 

A total of 82 patients were initially included in the study. Thirty three patients were excluded from the analyses due to either a too low number of measurements ($< 5$; $N = 4$), or a lack of variance in the response categories (smallest response category must contain at least $5\%$ of the responses; $N = 29$). This resulted in 49 patients that were included in the analyses. (Excluded patients (mean age = $48.79$ years, SD = $14.09$ years, $72.73$\% women) missed on average $28.10$\% of the measurements, and completed on average $60.39$ measurements (SD = $30.33$). These patients were admitted between 1988 and 1994, and were admitted on average for $209.45$ days (SD = $119.59$ days, min = $53$ days, max = $536$ days)). Excluded patients completed significantly less measurements than included patients ($t(38.03) = 2.77, p = 0.009$). Included patients had a mean age of $47.92$ ($SD = 13.13$ years) at the time of admission to the closed ward, with $71.43$\% women. These patients missed on average $9.96$\% of the measurements, and registered on average $75.71$ measurements (SD = $11.29$). Patients were admitted between 1988 and 1994, and were admitted on average for $179.35$ days (SD = $129.75$ days, min = $49$ days, max = $572$ days). Non-parametric $t$-tests revealed that the excluded and included patients did not significantly differ in age ($t(66.46) = 0.28, p = 0.78$), and admission period ($t(63.45) = 1.012, p = 0.32$). Under the \emph{EU General Data Protection Regulation}, we are not allowed to publish raw results. Result figures for all patients can be found online \citep{OSFKosII}.

\begin{figure}
\includegraphics[width = 0.99\textwidth]{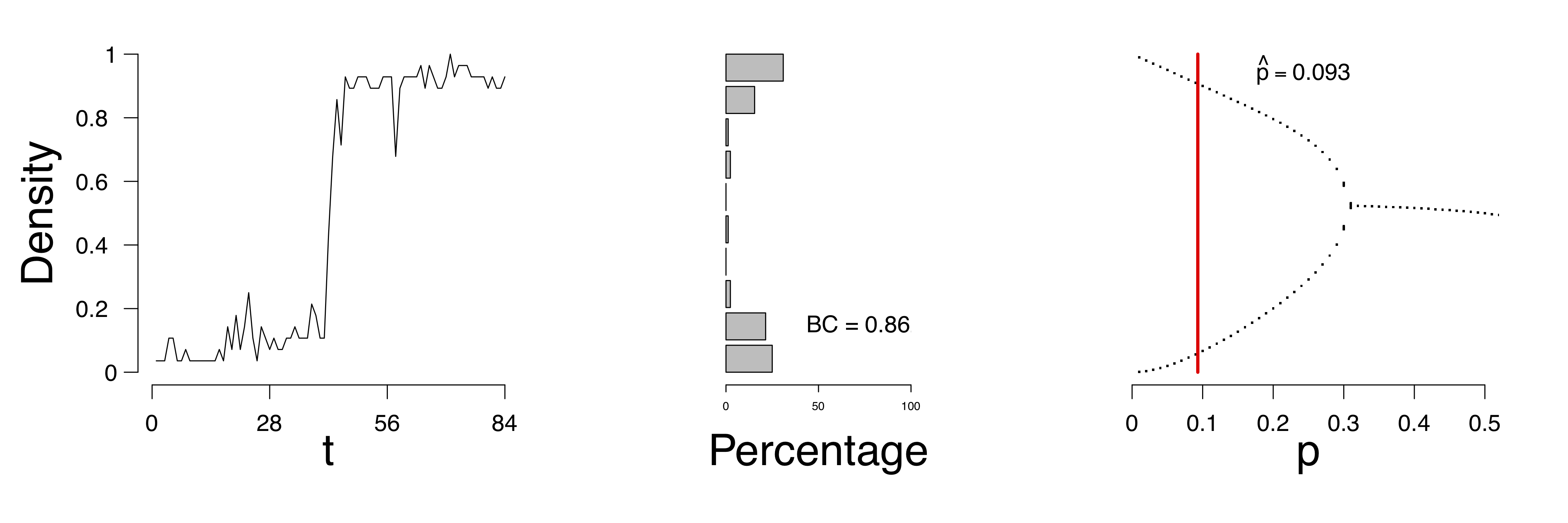}
\caption{Proportion of active symptoms (left) and bifurcation diagram (right) of one participant from the the Groningen data. BC = bimodality coefficient. Red line indicates the estimate $\hat{p}$.}
\label{fig:panel_groningen}
\end{figure}

Figure \ref{fig:panel_groningen} shows the evolution of the density (left figure), a distribution of the density $\rho_t$ (frequency of the number of active nodes; middle figure), and the estimate of $\hat{p}$ in the bifurcation diagram (right figure) of a single patient. Figures of all patients are available online. According to the mean field model $87.8$\% of the patients had an expectancy for a transition. This is not surprising given that the sample is from a population of patients in a psychiatric ward. To compare, we calculated the bimodality coefficient (BC), which uses a function of the skewness and kurtosis from the distribution of the time series of the proportion of symptoms (see \citet{Hosenfeld2015} for details). The BC classified $59.2\%$ of the cases as being bimodal. When we compare the results from the MFA to the BC, we see that the methods agree in $55.1$\% of the cases. In the case of the patient, whose results are depicted in Figure \ref{fig:panel_groningen}, the BC is very high ($0.86$), which is reflected in the shape of the distribution of the density and corresponds to the result of the MFA.

We investigated the robustness of the mean field model in an empirical setting by running a subset analysis. This analysis is similar to the one we conducted with simulated data that is described earlier. We randomly selected either $50\%$ or $75\%$ of the time points per patient and used ML estimation to estimate $\hat{p}$. Results showed that in $96.3\%$ of the participants, taking a subset of the data resulted in the same conclusion according to the mean field model. For the BC, we found that, taking a subset of the data resulted in the same conclusion in $86.6\%$ of the patients. This shows that the mean field model is fairly robust when one does not use all the data available.

\subsection{Example 2: General sample}

Participants were originally recruited in a nation-wide study called HoeGekIsNL (in English: HowNutsAreTheDutch) and have been described in detail in a previous paper \citep{VanderKrieke2015}. Participants in this study filled out a 43-item questionnaire that consisted of new items created for this study, and items from existing and validated questionnaires. Participants completed this questionnaire three times a day with a six-hour interval between the time points, for a period of 31 days, resulting in a maximum of 93 measurements per participant \citep{VanderKrieke2015}. 


\begin{table}
\centering
\begin{tabular}{p{1cm}p{5cm}p{6cm}}
 \hline
	Item & Meaning & Range \\ 
 \hline
1 & I feel relaxed & not at all (0) -- very much (100)\\
2 & I feel gloomy & not at all (0) -- very much (100)\\
3 & I feel energetic & not at all (0) -- very much (100)\\
4 & I feel anxious & not at all (0) -- very much (100)\\
5 & I feel enthusiastic & not at all (0) -- very much (100)\\
6 & I feel nervous & not at all (0) -- very much (100)\\
7 & I feel content & not at all (0) -- very much (100)\\
8 & I feel irritable & not at all (0) -- very much (100)\\
9 & I feel calm & not at all (0) -- very much (100)\\
10 & I feel dull & not at all (0) -- very much (100)\\
11 & I feel cheerful & not at all (0) -- very much (100)\\
12 & I feel tired & not at all (0) -- very much (100)\\
13 & I feel valued & not at all (0) -- very much (100)\\
14 & I feel lonely & not at all (0) -- very much (100)\\
15 & I feel I fall short & not at all (0) -- very much (100)\\
16 & I feel confident & not at all (0) -- very much (100)\\
17 & I worry a lot & not at all (0) -- very much (100)\\
18 & I am easily distracted & not at all (0) -- very much (100)\\
19 & I feel my life is worth living & not at all (0) -- very much (100)\\
20 & I am unbalanced & not at all (0) -- very much (100)\\
21 & I am in the here and now & not at all (0) -- very much (100)\\
22 & My appetite is.. & much small than usual (0) -- much larger than usual (100)\\
23 & Since the last measurement I had a laugh & not at all (0) -- very much (100)\\
 \hline
\end{tabular}
\caption{Items that were included in the analysis, the meaning of each item, and the response range in word and number.}
\label{tab:NL_labels}
\end{table}

From the original questionnaire, we selected items that pertained to mood states (21 items), appetite (one item) and laughter (one item), ending up with 23 items. Table \ref{tab:NL_labels} shows a detailed description of the included items. We recoded 10 positive items so that high scores indicate a more negative affect on all items. All included items were measured on a 0-100 scale. We dichotomized the data using a median split. This means that we calculated the median for each item for each participant, and split the data accordingly. We coded all the responses below the median as `0', and everything above the median as `1'. We also considered using a $k$-means clustering to dichotomize the data, but as these results were very similar to the results that we present, we chose not to include these results here. We replaced any missing measurements with the previous measurement. We also considered removing the missing measurements entirely, but as we found nearly identical results, we chose not to present these results.

A total of 974 participants participated in this study. We excluded 182 participants from the analyses due to a too low number of measurements ($< 5$), resulting in 792 participants that were included in the remainder of this section. (Excluded participants (mean age = $41.17$ years, SD = $13.56$ years, $84.06$\% women) missed on average $88.57$\% of the measurements, and completed an average of $1.38$ measurements (SD = $1.35$)). Excluded participants completed significantly less measurements than included participants ($t(800.36) = -44.19, p < 0.001$). Included participants had a mean age of $40.21$ (SD = $13.48$ years) at the start of the data collection, with $82.49$\% women. These participants missed on average $35.81$\% of the measurements and registered on average $58.67$ measurements (SD = $36.37$). Non parametric $t$-tests revealed that excluded and included participants did not significantly differ in age ($t(269.42) = 0.86, p = 0.39$). We also looked at the mean scores of the \emph{Depression and Anxiety Stress Scale} \citep[DASS;][]{Lovibond1995a, Lovibond1995}, the \emph{Quick Inventory of Depressive Symptomatology} \citep[QIDS;][]{Rush2003, Rush2006}, and the \emph{Positive Affect Negative Affect Scale} \citep[PANAS;][]{Peeters1996, Raes2009}. Non parametric $t$-tests revealed that excluded and included participants did not significantly differ on the DASS ($t(94.039) = 1.59, p = 0.12$), the QIDS ($t(252.17) = 1.91, p = 0.057$), the positive items of the PANAS ($t(251.07) = -1.09, p = 0.27$) or the negative items of the PANAS ($t(241.61) = 1.67, p = 0.10$). Under the \emph{EU General Data Protection Regulation}, we are not allowed to publish raw results. Result figures for all participants can be found online \citep{OSFKosII}.

\begin{figure}
\includegraphics[width = 0.99\textwidth]{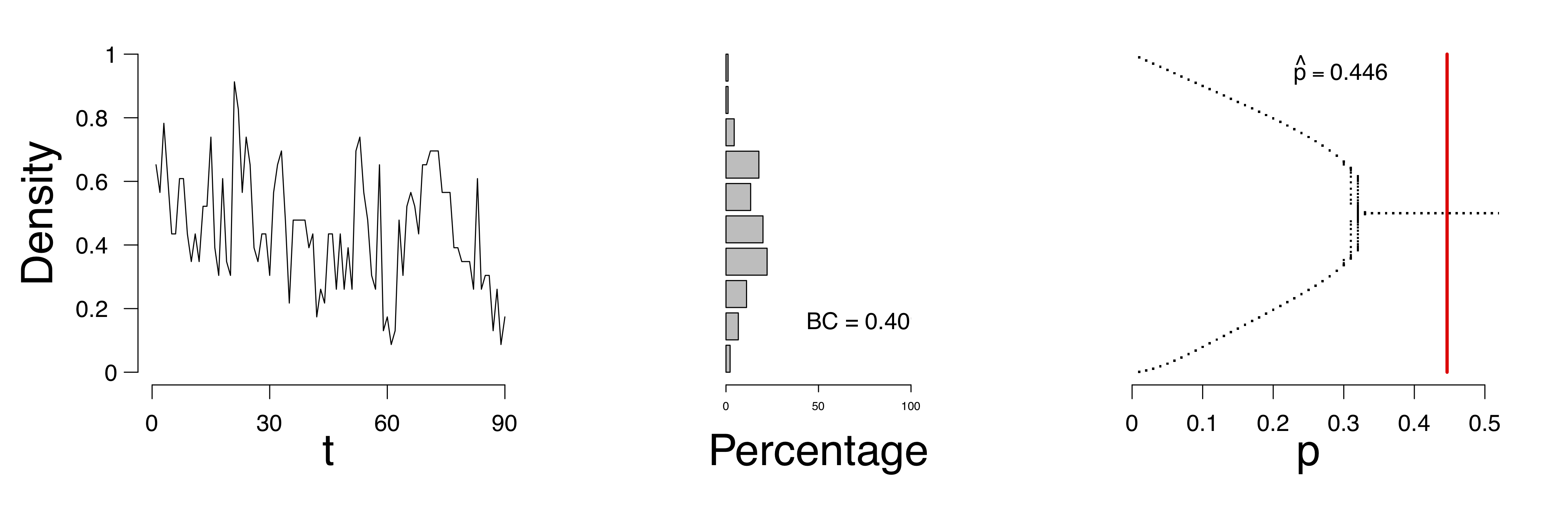}
\caption{Proportion of active symptoms (left) and bifurcation diagram (right) of one participant from the HowNutsAreTheDutch data. Red line indicates the estimate $\hat{p}$.}
\label{fig:panel_HGINL}
\end{figure}

Figure \ref{fig:panel_HGINL} shows the evolution of the density (left figure), a distribution of the density (frequency of the number of active nodes; middle figure), and the estimate of $\hat{p}$ in the bifurcation diagram (right figure) of a single participant. Figures of all participants are available online. According to the mean field model $20.8$\% of the participants have an expectancy for a transition. This is not surprising given that the sample is from the general population. To compare, we calculated the bimodality coefficient (BC), which uses a function of the skewness and kurtosis from the distribution of the time series of the proportion of symptoms (see \citet{Hosenfeld2015} for more details). The BC classified $31.9$\% of the participants as being bimodal. When we compare the results from the MFA to the BC, we see that the methods agree in $61.1$\% of the cases. In the case of the participant whose results are depicted in Figure \ref{fig:panel_HGINL}, the BC is not that high ($0.409$); this is reflected in the shape of the distribution of the density, which has a unimodal shape. This corresponds to the MFA result which indicates stability.

We investigated the robustness of the mean field model in an empirical setting by running a subset analysis. This analysis is similar to the one we conducted with simulated data that is described earlier. We randomly selected either $50\%$ or $75\%$ of the time points per participant and used ML estimation to estimate $\hat{p}$. Results showed that in $85.5\%$ of the participants, taking a subset of the data resulted in the same conclusion according to the mean field model. For the BC, we found that, taking a subset of the data resulted in the same conclusion in $81.6\%$ of the participants. This shows that the mean field model is fairly robust when one does not use all the data available.

%
\section{Discussion}

The present study combined dynamical systems theory and network theory to assess the expectancy for a transition, a sudden jump between two stable mood states, using a mean field model. We provided a simulation study and a validation study that both show that a mean field model can accurately identify individuals who may expect to experience a transition. We then applied the mean field model to two different empirical examples: data from patients admitted to a closed ward, and data from a general sample from a nation-wide study. Results from these applications show how our proposed method works in practice.

Results showed that the majority of the clinical sample could expect a transition. This indicates that these patients are likely to transition from a stable depressed mood state to a stable healthy mood state. Although we do not have any follow-up measures to investigate whether or not these transition actually occurred, we do know that these patients were eventually released from the closed ward. This may be an indication of a transition occurring in these patients. At the same time, the majority of the general sample were not expected to transition. This indicates that these participants will probably not transition from the stable state that they are currently in. As the analyses that we ran are of a probabilistic nature, we cannot know whether or not participants actually experienced a transition. It would be interesting for future research to run the HowNutsAreTheDutch study again for those participants that could expect a transition, and to investigate whether or not they received a clinical diagnosis.

When collecting time-series data, participants are requested multiple times a day to fill out a questionnaire for a certain period. This type of data collection demands time and energy of the participants. It thus makes sense that participants sometimes forget to complete a questionnaire, or are simply not up for it at that specific moment, for whatever reason. In the data that we analysed, we came across different ratios of missing data and completed measurements, ranging from no missing measurements to almost as much as $90$\%. Since we assumed a Markov model and so, the item responses should not change much and thus, we replaced missing measurements with the previous measurement. Adopting this approach for handling missing data decreases the variance that individual items may have, thereby increasing the probability that a participant may be expected to experience a transition. Although we did not find evidence that our analysis differed much if we removed these measurements altogether, at this point in time, there is no clear picture of the effects of missing data in the current analysis. Future research should focus on mapping the effects that different types of missing data have on the current analysis, and what the effect of various imputation methods have on the analyses.

The current study only allows for binary and no missing data. We applied different techniques for dichotomizing the data and handling missing data. Even though these different approaches did not lead to different conclusions, the current approach may not be ideal. Data are often imperfect: low variance within item scores, as well as missing data occurs recurrently in time-series data. More importantly, it can be argued that MDD symptoms may not be binary, but categorical or even continuous. One can imagine that there exists a scale on which individual MDD symptoms lie. For example, two participants may experience insomnia (one of the MDD symptoms as listed in the Diagnostic and Statistical Manual of Mental Disorders \citep{dsmv}), but the severity of this symptom may differ greatly between individuals. In the future we aim to expand the mean field model so that it allows continuous data as well as items with low variance.

The mean field model that we used in this paper has three assumptions: (1) we assume that each node in a graph has the same neighbourhood size, (2) nodes can only be in one of two states (active/inactive) and (3), we assume that all nodes in a graph show equal behavior. \citet{waldorp2019} showed one can deviate from the first assumption, whilst maintaining a high accuracy in estimating the probability $p$. We discussed the second assumption in more depth previously, which leaves us with the final assumption of the mean field model. In the current study, we operationalized the third assumption by fixating the probability $p$ to be equal for all nodes in the graph. However, it is unlikely that all symptoms of psychological disorders like MDD behave in a similar manner. For example, some individuals can handle sleep deprivation better than others. In this case, the ``sleep problems'' node would less easily be activated in individuals that can handle sleep deprivation in comparison to individuals that cannot handle sleep deprivation that well. A possible extension of the mean field model as is used in this paper is to vary the probability value $p$, which appears in the majority rule, for every node in the graph. In the example of sleep deprivation, we could operationalize the sensitivity difference by using different values for $p$ between nodes.

In the current study, we estimated the probability $p$ per individual for the entire time-series. This means that $p$ cannot change between time points. One may wonder if this value is supposed to be static, or that it could change between time points. The advantage of a static probability value is that it is easy to estimate. However, a static probability value may not reflect an individual's expectancy for a transition accurately. By allowing the probability $p$ to vary over time, one could gain more insight into how an individual moves throughout time with respect to $p$. One possible method to accomplish this is to work with a moving window, in which one uses a window to select a snippet within the time series to estimate $p$, and let that window move throughout the time series. In this situation, we can estimate $p$ several times on different segments of the time series; the size of the window will determine how many values are estimated. In the future we hope to expand the mean field model and allow for the probability $p$ to vary.

In the validation section of our current study we only looked at values for $p$, $p(e)$ and $p(w)$ between $0.1$ and $0.5$. Since these parameters are probabilities, their theoretical range lies between $0$ and $1$. Although we did run the validation study for values up to $0.9$, we chose not to present them as values rarely occur in empirical data. Also, at higher values for $p(e)$ and $p(w)$, the clustering within the network structures increases and can create some strange behaviours that are beyond the scope of this paper. A possible solution when dealing with high clustering values within a network is to switch to a so-called \emph{scale-free degree distribution}.

In conclusion, this study supports the notion that we are able to assess an individual's expectancy for a transition before it may occur. Based on the simulated and empirical examples provided here, we believe that the method is promising. We do emphasize that the predictions of our proposed model have not been verified using empirical evidence. We surely must investigate further to what extent the proposed method could be useful in clinical practice, but depending on the possible adjustments of the probability or majority rule in the model, the validity of the method could be high and therefore useful.


%

\section{Acknowledgements}

We thank Prof. Dr. Marieke Wichers and the HowNutsAreTheDutch project for letting us use their data. We also want to express out gratitude to Peter Groot, all the participants of the HowNutsAreTheDutch project, and the patients of the closed ward. The authors thank Prof. Dr. Marieke Wichers, Pia Tio, Dr. Maarten Marsman and all reviewers for their valuable contributions to the manuscript.

\section{Author Contributions}
JK and LW jointly generated the idea for the study. JK programmed the study. MG collected the data of the clinical sample, and HR contributed to the data collection of the general sample. JK wrote the analysis code and analyzed the data, LW verified the accuracy of those analyses. JK wrote the first draft of the manuscript, and all authors critically edited it. All authors approved the final submitted version of the manuscript.

\section{Conflicts of Interest}

The author(s) declare that there were no conflicts of interest with respect to the authorship or the publication of this article.

\section{Funding}

JK is partly funded by the Research Priority Area Yield, part of the Research Institute of Child Development and Education, University of Amsterdam, the Netherlands, and by the European Research Council Consolidator Grant received by Prof. Denny Borsboom (grant no. 647209).

The single-case experiment was designed by Prof. Dr. Marieke Wichers and Prof. Dr. Peter Groot. This study was supported by an Aspasia grant (Marieke Wichers., NWO grant), and by the Brain Foundation of the Netherlands (Marieke Wichers., grant no. F2012(1)-03).

The work on the HowNutsAreTheDutch project was supported by the Netherlands Organization for Scientific Research (NWO-ZonMW), by a VICI grant entitled Deconstructing Depression (no: 91812607) received by Prof. Dr. Peter de Jonge. Part of the HowNutsAreTheDutch project was realized in collaboration with the Espria Academy.

\section{Prior versions}

A pre-print of this manuscript is published at https://arxiv.org/pdf/1807.04269.pdf

\bibliographystyle{elsarticle-harv} 

\begin{thebibliography}{49}
\expandafter\ifx\csname natexlab\endcsname\relax\def\natexlab#1{#1}\fi
\providecommand{\url}[1]{\texttt{#1}}
\providecommand{\href}[2]{#2}
\providecommand{\path}[1]{#1}
\providecommand{\DOIprefix}{doi:}
\providecommand{\ArXivprefix}{arXiv:}
\providecommand{\URLprefix}{URL: }
\providecommand{\Pubmedprefix}{pmid:}
\providecommand{\doi}[1]{\href{http://dx.doi.org/#1}{\path{#1}}}
\providecommand{\Pubmed}[1]{\href{pmid:#1}{\path{#1}}}
\providecommand{\bibinfo}[2]{#2}
\ifx\xfnm\relax \def\xfnm[#1]{\unskip,\space#1}\fi
\bibitem[{{American Psychiatric Association}(2013)}]{dsmv}
\bibinfo{author}{{American Psychiatric Association}}, \bibinfo{year}{2013}.
\newblock \bibinfo{title}{{Diagnostic and Statistical Manual of Mental
  Disorders (DSM-5)}}.
\newblock \bibinfo{edition}{5th} ed., \bibinfo{publisher}{Washington, DC:
  Author}.
\bibitem[{Balister et~al.(2006)Balister, Bollob{\'{a}}s and
  Kozma}]{Balister2006}
\bibinfo{author}{Balister, P.}, \bibinfo{author}{Bollob{\'{a}}s, B.},
  \bibinfo{author}{Kozma, R.}, \bibinfo{year}{2006}.
\newblock \bibinfo{title}{{Large deviations for mean field models of
  probabilistic cellular automata}}.
\newblock \bibinfo{journal}{Random Structures {\&} Algorithms}
  \bibinfo{volume}{29}, \bibinfo{pages}{399--415}.
\bibitem[{Bollob{\'{a}}s(2001)}]{bollobas2001}
\bibinfo{author}{Bollob{\'{a}}s, B.}, \bibinfo{year}{2001}.
\newblock \bibinfo{title}{{Random Graphs}}.
\newblock \bibinfo{publisher}{Cambridge University Press},
  \bibinfo{address}{Cambridge, UK}.
\bibitem[{Borsboom(2017)}]{Borsboom2017}
\bibinfo{author}{Borsboom, D.}, \bibinfo{year}{2017}.
\newblock \bibinfo{title}{{A network theory of mental disorders}}.
\newblock \bibinfo{journal}{World Psychiatry} \bibinfo{volume}{16},
  \bibinfo{pages}{5--13}.
\bibitem[{Borsboom et~al.(2011)Borsboom, Epskamp, Kievit, Cramer and
  Schmittmann}]{Borsboom2011}
\bibinfo{author}{Borsboom, D.}, \bibinfo{author}{Epskamp, S.},
  \bibinfo{author}{Kievit, R.A.}, \bibinfo{author}{Cramer, A.O.J.},
  \bibinfo{author}{Schmittmann, V.D.}, \bibinfo{year}{2011}.
\newblock \bibinfo{title}{{Transdiagnostic Networks Commentary on
  Nolen-Hoeksema and Watkins (2011)}}.
\newblock \bibinfo{journal}{Perspectives on Psychological Science}
  \bibinfo{volume}{6}, \bibinfo{pages}{610--614}.
\bibitem[{Cramer et~al.(2016)Cramer, van Borkulo, Giltay, van~der Maas,
  Kendler, Scheffer and Borsboom}]{Cramer2016}
\bibinfo{author}{Cramer, A.O.J.}, \bibinfo{author}{van Borkulo, C.D.},
  \bibinfo{author}{Giltay, E.J.}, \bibinfo{author}{van~der Maas, H.L.J.},
  \bibinfo{author}{Kendler, K.S.}, \bibinfo{author}{Scheffer, M.},
  \bibinfo{author}{Borsboom, D.}, \bibinfo{year}{2016}.
\newblock \bibinfo{title}{{Major depression as a complex dynamical system}}.
\newblock \bibinfo{journal}{PloS ONE} \bibinfo{volume}{11},
  \bibinfo{pages}{e0167490}.
\bibitem[{Cramer et~al.(2012)Cramer, Sluis, Noordhof, Wichers, Geschwind,
  Aggen, Kendler and Borsboom}]{cramer2012}
\bibinfo{author}{Cramer, A.O.J.}, \bibinfo{author}{Sluis, S.},
  \bibinfo{author}{Noordhof, A.}, \bibinfo{author}{Wichers, M.},
  \bibinfo{author}{Geschwind, N.}, \bibinfo{author}{Aggen, S.H.},
  \bibinfo{author}{Kendler, K.S.}, \bibinfo{author}{Borsboom, D.},
  \bibinfo{year}{2012}.
\newblock \bibinfo{title}{{Dimensions of normal personality as networks in
  search of equilibrium: You can't like parties if you don't like people}}.
\newblock \bibinfo{journal}{European Journal of Personality}
  \bibinfo{volume}{26}, \bibinfo{pages}{414--431}.
\bibitem[{Csikszentmihalyi and Larson(1987)}]{csikszentmihalyi1987}
\bibinfo{author}{Csikszentmihalyi, M.}, \bibinfo{author}{Larson, R.},
  \bibinfo{year}{1987}.
\newblock \bibinfo{title}{{Validity and reliability of the experience-sampling
  method}}.
\newblock \bibinfo{journal}{The Journal of Nervous and Mental Disease}
  \bibinfo{volume}{175}, \bibinfo{pages}{526--536}.
\bibitem[{Durett(2007)}]{durrett2007}
\bibinfo{author}{Durett, R.}, \bibinfo{year}{2007}.
\newblock \bibinfo{title}{{Random Graph Dynamics}}.
\newblock \bibinfo{publisher}{Cambridge, UK: Cambridge University Press}.
\bibitem[{Epskamp(2015)}]{IsingSampler}
\bibinfo{author}{Epskamp, S.}, \bibinfo{year}{2015}.
\newblock \bibinfo{title}{{IsingSampler: Sampling Methods and Distribution
  Functions for the Ising Model}}.
\newblock \URLprefix \url{https://cran.r-project.org/package=IsingSampler}.
\bibitem[{Fleming and Harrington(1978)}]{Fleming1978}
\bibinfo{author}{Fleming, T.R.}, \bibinfo{author}{Harrington, D.P.},
  \bibinfo{year}{1978}.
\newblock \bibinfo{title}{{Estimation for discrete time nonhomogeneous Markov
  chains}}.
\newblock \bibinfo{journal}{Stochastic Processes and their Applications}
  \bibinfo{volume}{7}, \bibinfo{pages}{131--139}.
\bibitem[{Golubitsky and Stewart(2003)}]{Golubitsky2003}
\bibinfo{author}{Golubitsky, M.}, \bibinfo{author}{Stewart, I.},
  \bibinfo{year}{2003}.
\newblock \bibinfo{title}{{The symmetry perspective: from equilibrium to chaos
  in phase space and physical space}}.
\newblock \bibinfo{publisher}{Springer Science {\&} Business Media}.
\bibitem[{Gordijn et~al.(1998)Gordijn, Beersma, Bouhuys and van~den
  Hoofdakker}]{Gordijn1998}
\bibinfo{author}{Gordijn, M.}, \bibinfo{author}{Beersma, D.},
  \bibinfo{author}{Bouhuys, A.}, \bibinfo{author}{van~den Hoofdakker, R.},
  \bibinfo{year}{1998}.
\newblock \bibinfo{title}{{Mood variability and sleep deprivation effect as
  predictors of therapeutic response in depression}}.
\newblock \bibinfo{journal}{Sleep-wake research in the Netherlands}
  \bibinfo{volume}{9}, \bibinfo{pages}{41--44}.
\bibitem[{Gordijn et~al.(1994)Gordijn, Beersma, Bouhuys, Reinink and van~den
  Hoofdakker}]{Gordijn1994}
\bibinfo{author}{Gordijn, M.}, \bibinfo{author}{Beersma, D.},
  \bibinfo{author}{Bouhuys, A.}, \bibinfo{author}{Reinink, E.},
  \bibinfo{author}{van~den Hoofdakker, R.}, \bibinfo{year}{1994}.
\newblock \bibinfo{title}{{A longitudinal study of diurnal mood variation in
  depression: characteristics and significance}}.
\newblock \bibinfo{journal}{Journal of Affective Disorders}
  \bibinfo{volume}{31}, \bibinfo{pages}{261--273}.
\bibitem[{Guloksuz et~al.(2017)Guloksuz, Pries and {Van Os}}]{Guloksuz:2017}
\bibinfo{author}{Guloksuz, S.}, \bibinfo{author}{Pries, L.K.},
  \bibinfo{author}{{Van Os}, J.}, \bibinfo{year}{2017}.
\newblock \bibinfo{title}{{Application of network methods for understanding
  mental disorders: pitfalls and promise}}.
\newblock \bibinfo{journal}{Psychological medicine} \bibinfo{volume}{47},
  \bibinfo{pages}{2743--2752}.
\bibitem[{Guly{\'{a}}s et~al.(2013)Guly{\'{a}}s, Kampis and
  Legendi}]{gulyas2013}
\bibinfo{author}{Guly{\'{a}}s, L.}, \bibinfo{author}{Kampis, G.},
  \bibinfo{author}{Legendi, R.O.}, \bibinfo{year}{2013}.
\newblock \bibinfo{title}{{Elementary models of dynamic networks}}.
\newblock \bibinfo{journal}{The European Physical Journal Special Topics}
  \bibinfo{volume}{222}, \bibinfo{pages}{1311--1333}.
\bibitem[{Hasselblatt and Katok(2003)}]{Hasselblatt2003}
\bibinfo{author}{Hasselblatt, B.}, \bibinfo{author}{Katok, A.},
  \bibinfo{year}{2003}.
\newblock \bibinfo{title}{{A First Course in Dynamics}}.
\newblock \bibinfo{publisher}{Cambridge University Press}.
\bibitem[{Hirsch et~al.(2004)Hirsch, Smale and Devaney}]{hirsch2012}
\bibinfo{author}{Hirsch, M.W.}, \bibinfo{author}{Smale, S.},
  \bibinfo{author}{Devaney, R.L.}, \bibinfo{year}{2004}.
\newblock \bibinfo{title}{{Differential equations, dynamical systems, and an
  introduction to chaos}}.
\newblock \bibinfo{publisher}{Oxford, UK: Academic press}.
\bibitem[{Holmgren(1996)}]{holmgren2012}
\bibinfo{author}{Holmgren, R.}, \bibinfo{year}{1996}.
\newblock \bibinfo{title}{{A first course in discrete dynamical systems}}.
\newblock \bibinfo{publisher}{Springer Science {\&} Business Media}.
\bibitem[{Hosenfeld et~al.(2015)Hosenfeld, Bos, Wardenaar, Conradi, van~der
  Maas, Visser and de~Jonge}]{Hosenfeld2015}
\bibinfo{author}{Hosenfeld, B.}, \bibinfo{author}{Bos, E.H.},
  \bibinfo{author}{Wardenaar, K.J.}, \bibinfo{author}{Conradi, H.J.},
  \bibinfo{author}{van~der Maas, H.L.J.}, \bibinfo{author}{Visser, I.},
  \bibinfo{author}{de~Jonge, P.}, \bibinfo{year}{2015}.
\newblock \bibinfo{title}{{Major depressive disorder as a nonlinear dynamic
  system: bimodality in the frequency distribution of depressive symptoms over
  time}}.
\newblock \bibinfo{journal}{BMC Psychiatry} \bibinfo{volume}{15},
  \bibinfo{pages}{1--9}.
\bibitem[{Kossakowski(2018)}]{OSFKosII}
\bibinfo{author}{Kossakowski, J.J.}, \bibinfo{year}{2018}.
\newblock \bibinfo{title}{{Results from "Combining Dynamical Systems Theory and
  Network Theory in Major Depressive Disorder"}}.
\newblock \bibinfo{howpublished}{Retrieved from https://osf.io/edyzp/}.
\bibitem[{Kossakowski and Cramer(2018)}]{Kossakowski2018}
\bibinfo{author}{Kossakowski, J.J.}, \bibinfo{author}{Cramer, A.O.J.},
  \bibinfo{year}{2018}.
\newblock \bibinfo{title}{{Complexity, chaos and catastrophe: Modeling
  psychopathology as a dynamic system}}, in: \bibinfo{booktitle}{Frontiers of
  Cognitive Psychology}. chapter \bibinfo{chapter}{Network Sc}.
\bibitem[{Kossakowski et~al.(2017)Kossakowski, Groot, Haslbeck, Borsboom and
  Wichers}]{Kossakowski2017}
\bibinfo{author}{Kossakowski, J.J.}, \bibinfo{author}{Groot, P.C.},
  \bibinfo{author}{Haslbeck, J.M.B.}, \bibinfo{author}{Borsboom, D.},
  \bibinfo{author}{Wichers, M.}, \bibinfo{year}{2017}.
\newblock \bibinfo{title}{{Data from `Critical Slowing Down as a Personalized
  Early Warning Signal for Depression'}}.
\newblock \bibinfo{journal}{Journal of Open Psychology Data}
  \bibinfo{volume}{5}, \bibinfo{pages}{1--3}.
\bibitem[{Kozma et~al.(2004)Kozma, Puljic, Balister, Bollob{\'{a}}s and
  Freeman}]{kozma2004}
\bibinfo{author}{Kozma, R.}, \bibinfo{author}{Puljic, M.},
  \bibinfo{author}{Balister, P.}, \bibinfo{author}{Bollob{\'{a}}s, B.},
  \bibinfo{author}{Freeman, W.J.}, \bibinfo{year}{2004}.
\newblock \bibinfo{title}{{Neuropercolation: a random cellular automata
  approach to spatio-temporal neurodynamics}}.
\newblock \bibinfo{journal}{Conference on Cellular Automata} ,
  \bibinfo{pages}{435--443}.
\bibitem[{Kozma et~al.(2005)Kozma, Puljic, Balister, Bollob{\'{a}}s and
  Freeman}]{kozma2005}
\bibinfo{author}{Kozma, R.}, \bibinfo{author}{Puljic, M.},
  \bibinfo{author}{Balister, P.}, \bibinfo{author}{Bollob{\'{a}}s, B.},
  \bibinfo{author}{Freeman, W.J.}, \bibinfo{year}{2005}.
\newblock \bibinfo{title}{{Phase transitions in the neuropercolation model of
  neural populations with mixed local and non-local interactions}}.
\newblock \bibinfo{journal}{Biological Cybernetics} \bibinfo{volume}{92},
  \bibinfo{pages}{367--379}.
\bibitem[{van~der Krieke et~al.(2015)van~der Krieke, Jeronimus, Blaauw,
  Wanders, Emerencia, Schenk, de~Vos, Snippe, Wichers, Wigman, Bos, Wardenaar
  and de~Jonge}]{VanderKrieke2015}
\bibinfo{author}{van~der Krieke, L.}, \bibinfo{author}{Jeronimus, B.F.},
  \bibinfo{author}{Blaauw, F.J.}, \bibinfo{author}{Wanders, R.B.K.},
  \bibinfo{author}{Emerencia, A.C.}, \bibinfo{author}{Schenk, H.M.},
  \bibinfo{author}{de~Vos, S.}, \bibinfo{author}{Snippe, E.},
  \bibinfo{author}{Wichers, M.}, \bibinfo{author}{Wigman, J.T.W.},
  \bibinfo{author}{Bos, E.H.}, \bibinfo{author}{Wardenaar, K.J.},
  \bibinfo{author}{de~Jonge, P.}, \bibinfo{year}{2015}.
\newblock \bibinfo{title}{{HowNutsAreTheDutch (HoeGekIsNL): A crowdsourcing
  study of mental symptoms and strengths}}.
\newblock \bibinfo{journal}{International Journal of Methods in Psychiatric
  Research} \bibinfo{volume}{25}, \bibinfo{pages}{123--144}.
\bibitem[{Kuznetsov(2013)}]{kuznetsov2013}
\bibinfo{author}{Kuznetsov, Y.A.}, \bibinfo{year}{2013}.
\newblock \bibinfo{title}{{Elements of applied bifurcation theory}}.
\newblock \bibinfo{publisher}{New York, USA: Springer-Verlag}.
\bibitem[{Lebowitz et~al.(1990)Lebowitz, Maes and Speer}]{Lebowitz1990}
\bibinfo{author}{Lebowitz, J.L.}, \bibinfo{author}{Maes, C.},
  \bibinfo{author}{Speer, E.R.}, \bibinfo{year}{1990}.
\newblock \bibinfo{title}{{Statistical mechanics of probabilistic cellular
  automata}}.
\newblock \bibinfo{journal}{Journal of Statistical Physics}
  \bibinfo{volume}{59}, \bibinfo{pages}{117--170}.
\bibitem[{van~de Leemput et~al.(2014)van~de Leemput, Wichers, Cramer, Borsboom,
  Tuerlinckx, Kuppens, van Nes, Viechtbauer, Giltay, Aggen, Derom, Jacobs,
  Kendler, van~der Maas, Neale, Peeters, Thiery, Zachar and
  Scheffer}]{leemput2014}
\bibinfo{author}{van~de Leemput, I.A.}, \bibinfo{author}{Wichers, M.},
  \bibinfo{author}{Cramer, A.O.J.}, \bibinfo{author}{Borsboom, D.},
  \bibinfo{author}{Tuerlinckx, F.}, \bibinfo{author}{Kuppens, P.},
  \bibinfo{author}{van Nes, E.H.}, \bibinfo{author}{Viechtbauer, W.},
  \bibinfo{author}{Giltay, E.J.}, \bibinfo{author}{Aggen, S.H.},
  \bibinfo{author}{Derom, C.}, \bibinfo{author}{Jacobs, N.},
  \bibinfo{author}{Kendler, K.S.}, \bibinfo{author}{van~der Maas, H.L.J.},
  \bibinfo{author}{Neale, M.C.}, \bibinfo{author}{Peeters, F.},
  \bibinfo{author}{Thiery, E.}, \bibinfo{author}{Zachar, P.},
  \bibinfo{author}{Scheffer, M.}, \bibinfo{year}{2014}.
\newblock \bibinfo{title}{{Critical Slowing Down as Early Warning for the Onset
  and Termination of Depression}}.
\newblock \bibinfo{journal}{Proceedings of the National Academy of Sciences}
  \bibinfo{volume}{111}, \bibinfo{pages}{87--92}.
\bibitem[{Lovibond and Lovibond(1995a)}]{Lovibond1995}
\bibinfo{author}{Lovibond, P.}, \bibinfo{author}{Lovibond, S.},
  \bibinfo{year}{1995}a.
\newblock \bibinfo{title}{{The structure of negative emotional states -
  comparison of the depression anxiety stress scale (DASS) with the beck
  depression and anxiety inventories}}.
\newblock \bibinfo{journal}{Behaviour Research and Therapy}
  \bibinfo{volume}{33}, \bibinfo{pages}{335--343}.
\bibitem[{Lovibond and Lovibond(1995b)}]{Lovibond1995a}
\bibinfo{author}{Lovibond, S.}, \bibinfo{author}{Lovibond, P.},
  \bibinfo{year}{1995}b.
\newblock \bibinfo{title}{{Manual for the depression anxiety stress scales}}.
\newblock \bibinfo{edition}{2nd editio} ed., \bibinfo{publisher}{Psychology
  Foundation}, \bibinfo{address}{Sydney, Australia}.
\bibitem[{Molenaar(2007)}]{Molenaar2007}
\bibinfo{author}{Molenaar, P.C.}, \bibinfo{year}{2007}.
\newblock \bibinfo{title}{{On the implications of classic ergodic theorems:
  Analysis of developmental processes has to focus on intra-individual
  variation}}.
\newblock \bibinfo{journal}{Developmental Psychobiology} \bibinfo{volume}{50},
  \bibinfo{pages}{60--69}.
\bibitem[{Newman and Watts(1999)}]{newman1999}
\bibinfo{author}{Newman, M.E.J.}, \bibinfo{author}{Watts, D.J.},
  \bibinfo{year}{1999}.
\newblock \bibinfo{title}{{Renormalization group analysis of the small-world
  network model}}.
\newblock \bibinfo{journal}{Physics Letters A} \bibinfo{volume}{263},
  \bibinfo{pages}{4--6}.
\bibitem[{O'Donnell(2014)}]{ODonnell:2014}
\bibinfo{author}{O'Donnell, R.}, \bibinfo{year}{2014}.
\newblock \bibinfo{title}{{Analysis of boolean functions}}.
\newblock \bibinfo{publisher}{Cambridge University Press}.
\bibitem[{Peeters et~al.(1996)Peeters, Ponds and Vermeeren}]{Peeters1996}
\bibinfo{author}{Peeters, F.}, \bibinfo{author}{Ponds, R.},
  \bibinfo{author}{Vermeeren, M.}, \bibinfo{year}{1996}.
\newblock \bibinfo{title}{{Affectiviteit en zelfbeoordeling van depressie en
  angst}}.
\newblock \bibinfo{journal}{Tijdschrift voor de Psychiatrie}
  \bibinfo{volume}{38}, \bibinfo{pages}{240--250}.
\bibitem[{{R Core Team}(2016)}]{RCoreTeam2016}
\bibinfo{author}{{R Core Team}}, \bibinfo{year}{2016}.
\newblock \bibinfo{title}{{R: A Language and Environment for Statistical
  Computing}}.
\newblock \URLprefix \url{https://www.r-project.org/}.
\bibitem[{Raes et~al.(2009)Raes, Daems, Feldman, Johnson and van
  Gucht}]{Raes2009}
\bibinfo{author}{Raes, F.}, \bibinfo{author}{Daems, K.},
  \bibinfo{author}{Feldman, G.}, \bibinfo{author}{Johnson, S.},
  \bibinfo{author}{van Gucht, D.}, \bibinfo{year}{2009}.
\newblock \bibinfo{title}{{A psychometric evaluation of the Dutch version of
  the responses to positive affect questionnaire}}.
\newblock \bibinfo{journal}{Psychologica Belgica} \bibinfo{volume}{49},
  \bibinfo{pages}{293--310}.
\bibitem[{Rajarshi(2012)}]{Rajarshi2012}
\bibinfo{author}{Rajarshi, M.B.}, \bibinfo{year}{2012}.
\newblock \bibinfo{title}{{Statistical Inference for Discrete Time Stochastic
  Processes}}.
\newblock \bibinfo{publisher}{New Delhi: Springer}.
\bibitem[{Rush et~al.(2006)Rush, Bernstein, Trivedi, Camody, Wisniewski, Mundt,
  Shores-Wilson, Biggs, Woo, Nierenberg and Fava}]{Rush2006}
\bibinfo{author}{Rush, A.}, \bibinfo{author}{Bernstein, L.},
  \bibinfo{author}{Trivedi, M.}, \bibinfo{author}{Camody, T.},
  \bibinfo{author}{Wisniewski, S.}, \bibinfo{author}{Mundt, J.},
  \bibinfo{author}{Shores-Wilson, K.}, \bibinfo{author}{Biggs, M.},
  \bibinfo{author}{Woo, A.}, \bibinfo{author}{Nierenberg, A.},
  \bibinfo{author}{Fava, M.}, \bibinfo{year}{2006}.
\newblock \bibinfo{title}{{An evaluation of the quick inventory of depressive
  symptomatology and the Hamilton rating scale for depression: a sequenced
  treatment alternatives to relieve depression trial report}}.
\newblock \bibinfo{journal}{Biological Psychiatry} \bibinfo{volume}{59},
  \bibinfo{pages}{493--501}.
\bibitem[{Rush et~al.(2003)Rush, Trivedi, Ibrahim, Camody, Arnow, Klein,
  Markowitz, Ninan, Kornstein, Manber, Thase, Kocsis and Keller}]{Rush2003}
\bibinfo{author}{Rush, A.}, \bibinfo{author}{Trivedi, M.},
  \bibinfo{author}{Ibrahim, H.}, \bibinfo{author}{Camody, T.},
  \bibinfo{author}{Arnow, B.}, \bibinfo{author}{Klein, D.},
  \bibinfo{author}{Markowitz, J.}, \bibinfo{author}{Ninan, P.},
  \bibinfo{author}{Kornstein, S.}, \bibinfo{author}{Manber, R.},
  \bibinfo{author}{Thase, M.}, \bibinfo{author}{Kocsis, J.},
  \bibinfo{author}{Keller, M.}, \bibinfo{year}{2003}.
\newblock \bibinfo{title}{{The 16-item quick inventory of depressive
  symptomatology (QIDS), clinician rating (QIDS-C), and self-report (QIDS-SR):
  a psychometric evaluation in patients with chronic major depression}}.
\newblock \bibinfo{journal}{Biological Psychiatry} \bibinfo{volume}{54},
  \bibinfo{pages}{573--583}.
\bibitem[{Sarkar(2000)}]{sarkar2000}
\bibinfo{author}{Sarkar, P.}, \bibinfo{year}{2000}.
\newblock \bibinfo{title}{{A brief history of cellular automata}}.
\newblock \bibinfo{journal}{ACM Computing Surveys (CSUR)} \bibinfo{volume}{32},
  \bibinfo{pages}{80--107}.
\bibitem[{Scheffer et~al.(2014)Scheffer, Bascompte, Brock, Brovkin, Carpenter,
  Dakos, Held, van Nes, Rietkerk and Sugihara}]{Scheffer2014}
\bibinfo{author}{Scheffer, M.}, \bibinfo{author}{Bascompte, J.},
  \bibinfo{author}{Brock, W.A.}, \bibinfo{author}{Brovkin, V.},
  \bibinfo{author}{Carpenter, S.R.}, \bibinfo{author}{Dakos, V.},
  \bibinfo{author}{Held, H.}, \bibinfo{author}{van Nes, E.H.},
  \bibinfo{author}{Rietkerk, M.}, \bibinfo{author}{Sugihara, G.},
  \bibinfo{year}{2014}.
\newblock \bibinfo{title}{{Early-warning signals for critical transitions}}.
\newblock \bibinfo{journal}{Nature} \bibinfo{volume}{46},
  \bibinfo{pages}{53--59}.
\bibitem[{Waldorp and Kossakowski(2019)}]{waldorp2019}
\bibinfo{author}{Waldorp, L.J.}, \bibinfo{author}{Kossakowski, J.J.},
  \bibinfo{year}{2019}.
\newblock \bibinfo{title}{{Mean field dynamics of stochastic cellular automata
  for random and small-world graphs}}.
\newblock \bibinfo{howpublished}{https://arxiv.org/pdf/1610.05105.pdf}.
\bibitem[{Watts and Strogatz(1998)}]{watts1998}
\bibinfo{author}{Watts, D.J.}, \bibinfo{author}{Strogatz, S.H.},
  \bibinfo{year}{1998}.
\newblock \bibinfo{title}{{Collective dynamics of ‘small-world' networks}}.
\newblock \bibinfo{journal}{Nature} \bibinfo{volume}{393},
  \bibinfo{pages}{440--442}.
\bibitem[{Wichers et~al.(2016)Wichers, Groot, Psychosystems, {ESM Group} and
  {ESW Group}}]{Wichers2016}
\bibinfo{author}{Wichers, M.}, \bibinfo{author}{Groot, P.C.},
  \bibinfo{author}{Psychosystems}, \bibinfo{author}{{ESM Group}},
  \bibinfo{author}{{ESW Group}}, \bibinfo{year}{2016}.
\newblock \bibinfo{title}{{Critical Slowing Down as a Personalized Early
  Warning Signal for Depression}}.
\newblock \bibinfo{journal}{Psychotherapy and Psychosomatics}
  \bibinfo{volume}{85}, \bibinfo{pages}{114--116}.
\bibitem[{Wit et~al.(2012)Wit, van~den Heuvel and Romeijn}]{Wit2012}
\bibinfo{author}{Wit, E.}, \bibinfo{author}{van~den Heuvel, E.},
  \bibinfo{author}{Romeijn, J.W.}, \bibinfo{year}{2012}.
\newblock \bibinfo{title}{{‘All models are wrong...': an introduction to
  model uncertainty}}.
\newblock \bibinfo{journal}{Statistica Neerlandica} \bibinfo{volume}{66},
  \bibinfo{pages}{217--236}.
\bibitem[{Wolfram(1984)}]{wolfram1984b}
\bibinfo{author}{Wolfram, S.}, \bibinfo{year}{1984}.
\newblock \bibinfo{title}{{Computation theory of cellular automata}}.
\newblock \bibinfo{journal}{Communications in Mathematical Physics}
  \bibinfo{volume}{96}, \bibinfo{pages}{15--57}.
\bibitem[{{World Health Organization}(2012)}]{WHO2012}
\bibinfo{author}{{World Health Organization}}, \bibinfo{year}{2012}.
\newblock \bibinfo{title}{{Depression, a hidden burden}}.
\newblock \bibinfo{howpublished}{Retrieved from: http://www.who.int}.
\bibitem[{von Zerssen(1986)}]{VonZerssen1986}
\bibinfo{author}{von Zerssen, D.}, \bibinfo{year}{1986}.
\newblock \bibinfo{title}{{Clinical self-rating scales (CSRS) of the Munich
  psychiatric information system (PSYCHIS M{\"{u}}nchen)}}, in:
  \bibinfo{booktitle}{Assessment of depression}. \bibinfo{publisher}{Springer},
  pp. \bibinfo{pages}{270--303}.

\end{thebibliography}

\end{document}